\documentclass[pre, superscriptaddress, aps, twocolumn, showpacs, longbibliography, footnoteinbib]{revtex4-2}

\usepackage{amsmath,amssymb,amsfonts}

\usepackage[T2A]{fontenc}
\usepackage[utf8]{inputenc}
\usepackage[english]{babel}
\usepackage{graphicx}
\usepackage{wrapfig}
\usepackage{hyperref}

\usepackage{color}
\newcommand{\red}[1]{{#1}}
\newcommand{\redd}[1]{{#1}}

\newcommand{\braket}[2]{\langle#1|#2\rangle}

\begin{document}

\title{Localization transition in non-Hermitian systems depending on reciprocity and hopping asymmetry}
\author{Daniil Kochergin}
\affiliation{Moscow Institute of Physics and Technology, Dolgoprudny 141700, Russia}
\affiliation{Laboratory of Complex Networks, Center for Neurophysics and Neuromorphic Technologies, Moscow, Russia}
\author{Vasilii Tiselko}
\affiliation{Laboratory of Complex Networks, Center for Neurophysics and Neuromorphic Technologies, Moscow, Russia}
\affiliation{Ioffe Institute of the Russian Academy of Sciences, Saint-Petersburg 194021, Russia}
\author{Arsenii Onuchin}
\affiliation{Laboratory of Complex Networks, Center for Neurophysics and Neuromorphic Technologies, Moscow, Russia}
\affiliation{Skolkovo Institute of Science and Technology, Moscow 121205, Russia}

\date{\today}

\begin{abstract}
    We studied the single-particle Anderson localization problem for non-Hermitian systems on directed graphs. \red{Random regular graph and} various undirected standard random graph models were modified by controlling reciprocity and hopping asymmetry parameters. We found the emergence of left, biorthogonal and right localized states depending on both parameters and graph structure properties such as node degree $d$. For directed random graphs, the occurrence of biorthogonal localization near exceptional points is described analytically and numerically. The clustering of localized states near the center of the spectrum and the corresponding mobility edge for left and right states are shown numerically. Structural features responsible for localization, such as topologically invariant nodes or drains and sources, were also described. Considering the diagonal disorder, we observed the disappearance of localization dependence on reciprocity around $W \sim 20$ for a random regular graph $d=4$. With a small diagonal disorder, the average biorthogonal fractal dimension drastically reduces. Around $W \sim 5$ localization scars occur within the spectrum, alternating as vertical bands of clustering of left and right localized states.
\end{abstract}

\maketitle

\section{Introduction}\label{sec:int}

Anderson localization (AL) is a fundamental phenomenon corresponding to a metal-insulator phase transition where localized states in a system arise due to on-site energy disorder. Single-particle AL has attracted much attention in the context of localization in many-particle interacting systems (MBL). It has been shown~\cite{altshuler1997quasiparticle} that phase transition to MBL phase can be seen as Anderson transition on Bethe-lattice~\cite{abou1973selfconsistent}. A similar behavior was described for the random regular graph model (RRG), which now serves as a toy-model of the Hilbert space of the many-body system problem~\cite{Tikhonov2021From}.

In addition to diagonal disorder, localized states can occur due to structural disorder, as has been shown for Wegner model~\cite{Wegner1979Disordered}, a system with only local interactions between spins and fermions~\cite{Smith2017Disorder}, models with long-range interaction (for example, Euclidean random matrices~\cite{Kutlin20220Renormalization,goetschy2013euclidean}), strong degree fluctuation with a heavy vertex~\cite{Biroli1999Asingle,PastorSatorras2015Distinct,Nechaev2017Path,matyushina2023statistics}, in exponential networks with chemical potential of $k$-cycles~\cite{Avetisov2019Localization,Valba2021Interacting,Kochergin2023Anatomy} or in partially disordered RRG \cite{Valba2022Mobility,Kochergin2023Robust}. The structural disorder can also be born in directed graphs with random direction distribution or asymmetric hopping which leads to a non-Hermitian system. 

Similarly, complex natural networks such as neural networks~\cite{brunel2000dynamics}, ecosystems~\cite{bascompte2009disentangling}, gene regulatory networks~\cite{milo2002network}, social networks~\cite{kwak2010twitter}, and the World Wide Web~\cite{broder2000graph}, can be represented as extensive networks with directed connections. The right eigenvectors of adjacency matrices in directed graphs are used in algorithms for determining node centrality \cite{bonacich1972factoring, restrepo2006characterizing, martin2014localization}, detecting communities \cite{krzakala2013spectral, bordenave2015non, kawamoto2018algorithmic}, matrix completion \cite{bordenave2022detection}, stochastic processes~\cite{Wasserman1980AStochastic,Tapias2022Localization}.

Localization in non-Hermitian physical systems is mainly studied in one-dimensional chains, for example, Hatano-Nelson model with asymmetric hopping~\cite{HN_PhysRevLett.77.570,HN_PhysRevB.56.8651,HN_PhysRevB.58.8384}. Although the states in the presence of diagonal disorder in one-dimensional systems are always localized, for example, as in Aubry-Andr\'e model with an aperiodic slowly varying potential, the mobility edge that separates localized and delocalized states exists in both Hermitian and non-Hermitian cases~\cite{DasSarma1988Mobility,DasSarma1990Localization,Liu2020Generalized}. The existence of localization was studied in Ginibre ensemble and other non-Hermitian random matrix models~\cite{peron2020spacing,sa2020complex,Hamazaki2019NonHermitian,Huang2020Anderson,Tzortzakakis2020NonHermitian,Luo2021Universality,DeTomasi2022NonHermitian,detomasi2023nonhermiticity}. The papers~\cite{Neri2012Spectra,Neri2016Eigenvalue,Metz2019Spectral} discuss the spectral properties of directed graphs and the dependence of the right eigenvector distribution for isolated eigenvalue and eigenvalues on the boundary of a continuous region of the spectrum for different distributions of weights and degrees of outgoing edges~\cite{Metz2021Localization}. 

Non-Hermitian systems can have drastic differences between periodic and open boundary conditions. In the open boundary conditions left and right eigenvectors can be localized on opposite edges of the system. This phenomenon is called non-Hermitian skin effect. Besides it, a non-Hermitian system can have exceptional degeneracy where eigenvalue and eigenvector collapse. These points are tightly bounded to localization on system edges~\cite{Kunst2018Biorthogonal,Yao2018Edge,Xiong2018Why,Kawabata2023Entanglement, Bergholtz2021Exceptional,Okuma2023NonHermitian}. 

In the present study, we discuss a transition to a localized phase in dependence on the graph's bidirectionality and the bandwidth between directions. The paper is organized as follows. In Section~\ref{sec:ex}, we demonstrate that localization occurs on directed random graph models around exceptional degeneracy, in Section~\ref{sec:model} we describe the model and the main methods used. In Section~\ref{sec:nr} we numerically study the presence of localization in \red{modified RRG and various undirected standard random graph models} and consider the topological reasons for the occurrence of localization using the example of a regular lattice with toroidal boundary conditions. Additionally, in Section~\ref{sec:dd}, we numerically consider combined disorder (structural+diagonal). Finally, in Section~\ref{sec:con} we summarize our results and raise some problems for future study.

\section{Existence of localization in directed graphs}\label{sec:ex}

In this section we demonstrate several examples of how exceptional degeneracy in non-Hermitian systems is associated with the existence of localization in directed random graphs.

For \red{any} non-Hermitian \red{matrix, $M$,} left and right eigenstates are not complex conjugated which is the reason why we use biorthogonal quantum mechanic~\cite{Brody2014Biorthogonal,Ashida2020NonHermitian} with $\langle \psi^L_i | M | \psi^R_j \rangle = \lambda_{i}$, $\langle \psi^L_i | \psi^R_j \rangle = \delta_{ij}$, where $\psi^{L(R)}_i$ is left (right) eigenstate for $\lambda_i$ of $M$. 
\red{Also, we use the notation $\phi^{L(R)}$ that denotes eigenvectors with the norm $||\phi^{L(R)}\rangle|^2=1$.}
\red{In this paper, except Section~\ref{sec:dd}, we focus on non-symmetric real adjacency matrices $H$.}

An example of a minimal model with exceptional degeneracy is two nodes connected by two oppositely directed edges with weights $1-\epsilon$ and $\epsilon$. \red{Equation for right eigenvector is
\begin{equation}
\begin{pmatrix}
        0 & 1-\epsilon \\
        \epsilon & 0 \\
    \end{pmatrix}
    \phi^R
    = \lambda
    \phi^R. 
\end{equation}
Solution of which is $\lambda=\pm\sqrt{\epsilon(1-\epsilon)}$, $\phi^L=(\pm\sqrt{\epsilon},\sqrt{1-\epsilon})$, $\phi^R=(\sqrt{1-\epsilon},\pm\sqrt{\epsilon})^T$.} At the exceptional point (EP), $\epsilon=0$, the adjacency matrix is defective, i.e. non-diagonalizable and has Jordan block form. In this case, the eigenvalues degenerate, while the left and right eigenvectors coalesce\red{, $\lambda=0$, $\phi^L=(0,1)$, $\phi^R=(1,0)$}. 

We consider a more general example: a directed graph that can be divided into two clusters $A$ and $B$ with links connecting nodes between clusters with weight $1-\epsilon$ in one direction, and a weight $\epsilon$ in the opposite. The edges are randomly distributed inside the blocks, while between \red{them from block $A$ to block $B$,} $A \rightarrow B$,  the arrow shows the directed edges connecting the blocks at EP. The equation for the right eigenstate in matrix form:
\begin{equation}\label{eq:H_AB}
    \red{H\phi^R=}
    \begin{pmatrix}
        A & y \\
        x & B \\
    \end{pmatrix}
    \begin{pmatrix}
        \phi^R_A \\
        \phi^R_B \\
    \end{pmatrix}
    = \lambda
    \begin{pmatrix}
        \phi^R_A \\
        \phi^R_B \\
    \end{pmatrix}. 
\end{equation}
where size of the $A$ block is $n \times n$, size of the $B$ block is $m \times m$, $n+m=N$, nonzero elements in $y$ are equal to $1-\epsilon$, non-zero elements in $x$ are equal to $\epsilon$, $x=\frac{\epsilon}{1-\epsilon}y^T$. 
\red{The capital letter subscript ($A$, $B$, etc.) stands for part of the eigenvector that corresponds to nodes in the block with the same letter.} The edges are fully directed between clusters at EP, which corresponds to $x=0$ in (\ref{eq:H_AB}). Then the energy levels are determined only by diagonal blocks: $\det(H-\lambda I)=\det(A-\lambda I)\det(B-\lambda I)=0$. $\phi^R_B=0$ if $\lambda^*$ satisfies $\det(A-\lambda^* I)=0$. In other words, the right eigenvector is \red{distributed} only over nodes from the block $A$. The distribution of the left eigenvector $\phi^L$ depends on the block $B$ structure. If $\det(B-\lambda^* I)\neq0$, then $\phi^L$ can have non-zero values in all graph nodes. Otherwise, $\phi^L_A=0$, which leads to $\braket{\phi^L}{\phi^R}=0$, and as a consequence the adjacency matrix becomes defective at EP.

Since \red{$\phi^L_B$ and} $\phi^R_A$ takes non-zero values at EP, \red{$\phi^L_A$ and} $\phi^R_B$ can be expressed as follows:
\begin{equation}
\begin{gathered}
    \phi^L_A=\phi^L_B x (\lambda I -A)^{-1}, \quad
    \phi^R_B=(\lambda I-B)^{-1} x \phi^R_A.\\
\end{gathered}
\end{equation}
If \red{the number of different orthogonal states (degree of degeneracy) of eigenvalue $\lambda^*$ are} $k_A$ and $k_B$ at corresponding blocks, then the normalized products of left and right eigenvectors:
\begin{equation}\label{eq:AB}
    \begin{gathered}
    \begin{aligned}
    \psi^L_A \psi^R_A &= Z^{-1}
    \phi^L_B x (\lambda-\lambda^*)^{-k_A} R'_A(\lambda^*) \phi^R_A \sim \\
    &\sim \frac{\mathcal{O} \left((\lambda-\lambda^*)^{k_B}\right)}
    {\mathcal{O} \left((\lambda-\lambda^*)^{k_A}\right)+\mathcal{O} \left((\lambda-\lambda^*)^{k_B}\right)}
    \end{aligned},\\
    \begin{aligned}
    \psi^L_B \psi^R_B &= Z^{-1}
    \phi^L_B R'_B(\lambda^*)(\lambda-\lambda^*)^{-k_B} x \phi^R_A \sim \\
    &\sim \frac{\mathcal{O} \left((\lambda-\lambda^*)^{k_A}\right)}
    {\mathcal{O} \left((\lambda-\lambda^*)^{k_A}\right)+\mathcal{O} \left((\lambda-\lambda^*)^{k_B}\right)}
    \end{aligned},\\
    Z=\phi^L_B R'_B(\lambda^*)(\lambda-\lambda^*)^{-k_B} x + x (\lambda-\lambda^*)^{-k_A} R'_A(\lambda^*) \phi^R_A,
    \end{gathered}
\end{equation}
where $(\lambda I-A)^{-1}=\operatorname{adj}(\lambda I-A)/\det(\lambda I-A)=(\lambda-\lambda^*)^{-k_A} R'_A(\lambda^*)$. In (\ref{eq:AB}) we consider the product of left and right eigenvectors of infinitesimal order $(\lambda-\lambda^*)$. The distribution of products of the left and right eigenvectors depends on the degree of \red{degeneraсy}. If degrees \red{of degeneraсy} are different, $k_A>k_B$ ($k_A<k_B$), then the product of left and right eigenvectors is \red{distributed} only on block $A$, $\psi^L_B \psi^R_B\to0$ ($B$, $\psi^L_A \psi^R_A\to0$), which we call biorthogonal localization. If they are identical, $k_A=k_B$, \red{then $\psi^L_A\psi^R_A$ and $\psi^L_B\psi^R_B$ have the same order of infinitesimals and} the state is \red{equally distributed on both blocks, i.e.} delocalized. In the limit $\epsilon\to0$, diagonal elements are normalized, but elements $\psi^L_B \psi^R_A \sim \left(\mathcal{O}\left((\lambda-\lambda^*)^{k_A} \epsilon\right)+\mathcal{O}\left(\epsilon(\lambda-\lambda^*)^{k_B}\right)\right)^{-1}$ tend to diverge.

It is necessary to clarify, that the eigenvalue after perturbation $(\lambda-\lambda^*) \sim \epsilon^{1/l}$ depends on the exceptional order of degeneracy that is equal to the size of the corresponding Jordan block $l$. The above Jordan block represents a one-dimensional chain of $l$ sites with edges, directed \red{from one boundary to opposite}, and with open boundary conditions. In this case for a $l$-length chain, perturbation around EP gives eigenvalue, $\lambda \approx \lambda^* + \epsilon^{1/l}\lambda_1 + \epsilon^{2/l}\lambda_2 + \mathcal{O}(\epsilon^{3/l})$~\cite{Bergholtz2021Exceptional}.

To show the existence of biorthogonal localization with separate localization of left and right eigenvectors we consider a more general case of the system with three blocks $A$, $B$ and $C$, $A \rightarrow B \rightarrow C$. Then the right eigenstate equation is
\begin{equation}\label{eq:ABС}
    \begin{pmatrix}
        A & y & 0 \\
        x & B & z \\
        0 & w & C \\
    \end{pmatrix} 
    \begin{pmatrix}
        \phi^R_A \\ \phi^R_B \\ \phi^R_C
    \end{pmatrix}
    =\lambda
    \begin{pmatrix}
        \phi^R_A \\ \phi^R_B \\ \phi^R_C
    \end{pmatrix},
\end{equation}
where $x$ and $w$ contain feedback edges (i.e. edges to opposite direction) and have an order of $\epsilon$. \red{If blocks $A$ and $C$ have at least one common eigenvalue $\lambda^*$ at EP ($x=0$, $w=0$) and $\lambda^*$ is not the eigenvalue of $B$, then the right (left) eigenvector is distributed only on block $A$ ($C$).} Since we know that $\phi^R_A$ is \red{non-zero} around EP, for the right eigenvector, we express the components of the blocks $B$ and $C$ through components of the block~$A$.
\begin{equation}
    \begin{aligned}
        \phi^R_B&=\left[\lambda I-B-z(\lambda I-C)^{-1}w\right]^{-1} x \phi^R_A ,\\
        \phi^R_C&=(\lambda I-C)^{-1} w \left[\lambda I-B-z(\lambda I -C)^{-1}w\right]^{-1} x \phi^R_A .\\
    \end{aligned}
\end{equation}
Similar formulas can be written for the left eigenvector with expressions through \red{$\phi^L_C$}. Let's compare the product of left and right eigenvectors of corresponding elements for different blocks by the order of the $\epsilon$ block:
\begin{equation}
    \begin{gathered}
    \phi^L_{A,C} \phi^R_{A,C} \sim 
    \frac{\mathcal{O}(\epsilon^2)}
    {\mathcal{O}(\epsilon^{k_{A,C}/l})+\mathcal{O}(\epsilon)}, \\
    \phi^L_{B} \phi^R_{B} \sim 
    \left(\frac{\mathcal{O}(\epsilon^{k_A/l+1})}
    {\mathcal{O}(\epsilon^{k_{A}/l})+\mathcal{O}(\epsilon)}\right) 
    \left(\frac{\mathcal{O}(\epsilon^{k_C/l+1})}
    {\mathcal{O}(\epsilon^{k_{C}/l})+\mathcal{O}(\epsilon)}\right).\\
    \end{gathered}
\end{equation}
Analogically to the two-block structure, we introduce the degree of degeneracy $k_{A,B,C}$ of the eigenvalue $\lambda^*$ for the corresponding blocks. $\phi^L_{B} \phi^R_{B}$ elements have a larger infinitesimal order, which means that they are suppressed compared to $\phi^L_{A,C} \phi^R_{A,C}$. In other words, there is biorthogonal localization on 
\red{both blocks $A$ and $C$ or on one of them} at small $\epsilon$ depending on \red{degree of degeneracy} and EP order.

Thus, we have shown the possibility of both the emergence or suppression of localization in the biorthogonal case with the presence of separate localization on the right and left eigenvectors.

\section{Model}\label{sec:model}

We consider a smooth transition from the undirected to the fully directed graph that varies the reciprocity parameter $r$ and the hopping asymmetry $p$. Hereinafter, the resulting graph will be called $rp$-network by the names of two control parameters.

\subsection{\texorpdfstring{$rp$}{rp}-network}
A traditional way to define network reciprocity is in terms of the ratio of bidirectional to unidirectional connections. Thus, for each network model we start from an undirected graph, and then the connections are modified as follows: taking the probability $r$ (reciprocity) we replace an undirected edge with two oppositely directed ones with weights $p$ and $1-p$ choosing a direction randomly. Otherwise, with the probability $1-r$ the undirected edge is changed to one directed in a random direction with weight of $1$. Therefore, the total bandwidth of the link between connected nodes is constant and equals to $1$. If $r=0$ or $p=0$ or $p=1$, the graph becomes an oriented directed graph. If $r=1$ and $p=0.5$, the graph is equivalent to undirected, where all edges have weights equal to $0.5$.

As we will show later, our model has a tendency to localization in a certain range of parameters. Nevertheless, due to the unidirectional nature of edges, the adjacency matrix often becomes defective and EP emerges. To avoid this issue we add feedback edges with small weight $\epsilon$ to unidirected edges and change the weight of the initial edge to $1-\epsilon$. \red{We call that procedure $\epsilon$-perturbation.}

For $rp$-network matrix element of Hamiltonian takes the form
\begin{equation}
    H_{nm}=t_{nm}A_{nm}
\end{equation}
All information about the weight and directions of the network edges is contained in $t_{nm}$. $A_{nm}$ is an element of the adjacency matrix of the initial undirected graph (i.e. undirected graph from which we have started forming $rp$-network) consisting of $1$ if there is an edge between $n$ and $m$ nodes and $0$ if the edge is absent.  

From Section~\ref{sec:ex} we know that localization may occur when one part of a graph is connected to another by edges pointing in the same direction. In the simplest case, one part of the graph consists of one vertex. For our model, we can estimate the critical value of the reciprocity $r_c$ for the random regular graph (RRG) with the degree of the vertex $d$, where localization occurs due to a single node with all incoming or outgoing edges. The probability of such a node occurring is $((1-r)/2)^d$. The other configurations have probabilities in power order $\gtrsim 2d\red{-1}$ and could be neglected \red{(see Sec.~\ref{sec:ool} for details)}. Since the localization on a node with all incoming or outgoing edges exists on left or right eigenvectors respectively, we consider localization on one of them. If we require that the graph has at least one such node, then $((1-r)/2)^d N \geq 1$, where $N$ is the number of nodes. Hence, the critical value of reciprocity is
\begin{equation}\label{eq:r_c}
    r_c=1-\frac{2}{N^{1/d}}.
\end{equation}
For RRG with $N=1024$ and $d=4$ the critical value is $r_c \approx 0.65$. We can also estimate the number of nodes with all in or out edges if $r=0$ then $n=N/2^d = 64$. Hence, for our $rp$-network, it is very common to have modes that are localized at least on the considered structural pattern with a corresponding eigenvalue equal to $0$. 

\red{From (\ref{eq:r_c}) we see that the critical value of reciprocity depends on $N$, and in the thermodynamic limit $r_c(N\to\infty)=1$, i.e., there is always at least one localized state.}

\subsection{Fractal dimensions}

To determine localized states, we use inverse partition ratio $IPR_q$ \red{\cite{Gong2018Topological,Xiao2022Topology,Suthar2022NonHermitian,DeTomasi2022NonHermitian}}:
\begin{equation}\label{eq:IPR(E)}
\begin{gathered}
IPR^S_{qi}=\frac{\sum_{n}^N |\psi^S_i(n)|^{2q}}{\left(\sum_{n}^N |\psi^S_i(n)|^{2}\right)^q},
\\
IPR^B_{qi}=\frac{\sum_n^N |\psi^L_i(n)\psi^R_i(n)|^q}
                    {\left(\sum_n^N |\psi^L_i(n)\psi^R_i(n)|\right)^q} ,
\end{gathered}
\end{equation}
where $IPR^S_q$ refers to separate eigenvectors, i.e., left and right; and $IPR^B_q$ to biorthogonal.
Because we distinguish left and right eigenstates, $IPR^L_i$, $IPR^R_i$, $IPR^B_i$ can have different values for the same state. \red{Physical systems require studying not only squared eigenvectors but also the product of left and right eigenvectors, because the physical properties of the system depend on both of them. For example, the density operator is $\rho=|\psi^R\rangle\langle\psi^L|$ or Green's function is $G(\lambda)=\sum_i \frac{|\psi^R_i\rangle\langle\psi^L_i|}{\lambda-\lambda_i}$~\cite{Metz2019Spectral}. Furthermore, in the context of the non-Hermitian skin effect, biorthogonal $IPR^B_q$ shows differences between skin-localized and bulk-localized states~\cite{Bergholtz2021Exceptional,Gong2018Topological,Xiao2022Topology,HN_PhysRevB.58.8384}.} 

The following expressions are valid for all types of $IPR_q$. Averaged over the spectrum:
\begin{equation}
    IPR_q=\frac{1}{N} \sum_i IPR_{qi}.
\end{equation}
As a measure of eigenfunction localization, the fractal dimension $D_q$ is considered:
\begin{equation}
    IPR_{q} \red{\sim}
    N^{-\tau_q}, \quad \tau_q=D_q(q-1),
\end{equation}
where $D_q\to0$ corresponds to a fully localized phase and $D_q\to1$ to a delocalized phase, when $N\to \infty$. If the value is intermediate, two options are available. First, the phase consists of non-ergodic extended states. Second, the phase is mixed. \red{Some of the states are ergodic; the others are localized. The parts can be divided by mobility edge, or there are localized scar-like states in the delocalized part of the spectrum, like in~\cite{Moudgalya2022Quantum,Kochergin2023Anatomy}.} In \red{this} case, the appearance of localized states can be revealed by high $IPR_q$ and fractal dimensions.
\begin{equation}\label{eq:IPRmixed}
    IPR_q \sim (1-\alpha) N^{-(q-1)} + \alpha \cdot 1,
\end{equation}
where $\alpha$ is a fraction of localized states. If there are any localized states, the first term \mbox{($\sim-(q-1)$)} is suppressed compared to the second term ($\sim-1$) in (\ref{eq:IPRmixed}). \red{For numerical calculations in the Section~\ref{sec:nr}} we use $q=4$ and $q=6$.

\red{To find the fractal dimension from $IPR_q$, we use $D_q=\log_2 IPR_q/\log_2 N$. Due to this definition, the fractal dimension remains dependent on the size $D_q(N)$~\red{\cite{Tikhonov2016Anderson}}. But since we seek the transition from localization to delocalization and not from ergodicity to non-ergodicity, we can neglect the dependence on $N$, which has extrapolation in the form $D_q(N)=D_q(\infty)+c_2/\redd{\log_2} N$~\red{\cite{Luca2014}}.} \redd{The last term has inverse logarithmic dependence; therefore, it is still relevant for calculating system size.}

\begin{figure*}[t]
    \centering
    \includegraphics[width=1.0\textwidth]{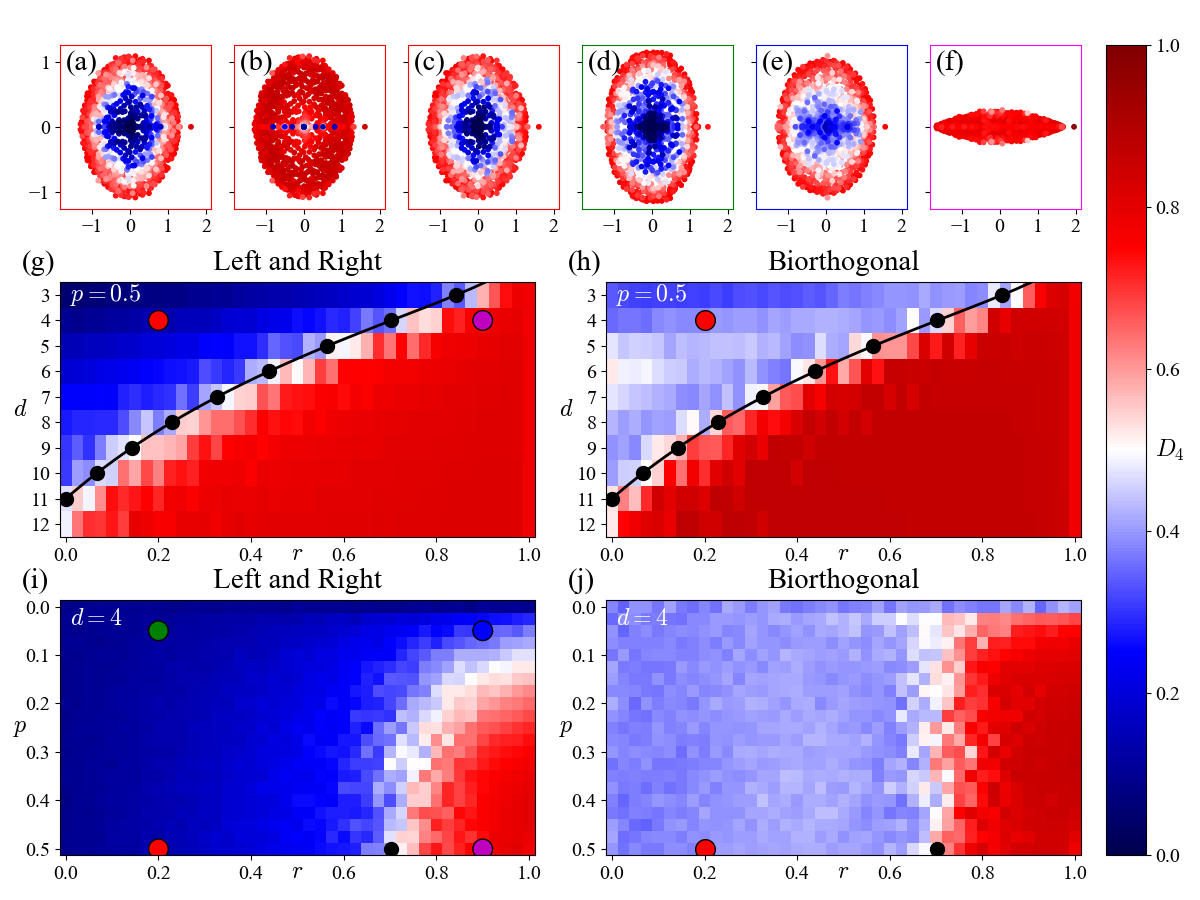}
    \caption{Numerical study of the single-particle Anderson transition for RRG using the fractal dimension $D_4$ as a localization measure. The results of calculations for networks with \mbox{$N=1024$} nodes are shown. (a)-(f)~Adjacency matrix spectra for different values of reciprocity $r$ and hopping asymmetry $p$. Eigenvalues are colored according to the values of the fractal dimension $D_4$, the color of the frame reflects the parameters set at the corresponding point in (g)-(j). (a)-(c) Same spectrum but colored by left, biorthogonal and right $D_4$. The left and right localized states cluster in the center of the adjacency matrix spectrum with the formation of a characteristic mobility edge. Dependence of the left and right~(g),(i) and biorthogonal~(h),(j) fractal dimension $D_4$ on reciprocity $r$ and node degree $d$ for $p=0.5$~(g),(h), $r$ and hopping asymmetry $p$ for $d=4$~(i),(j). Fractal dimensions are averaged over $16$ realizations, averaging over the left and right fractal dimensions calculated as independent. The black lines on~(g)\red{,(h) and the black dotes on~(i),}(j) denote the critical value of reciprocity calculated by (\ref{eq:r_c}) \red{for studied system size}.}  
    \label{fig:RRG_D4}
\end{figure*}

\begin{figure*}
\centering
    \includegraphics[width=1.0\textwidth]{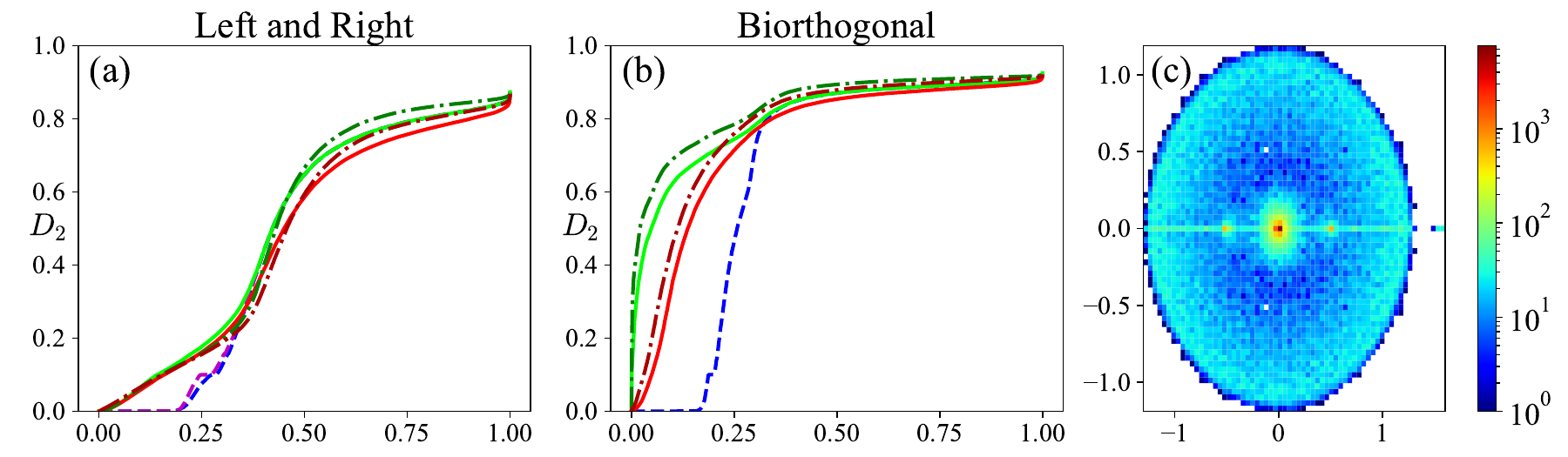}
    \caption{\red{Fractal dimension for states sorted in ascending order. (a)~Left and right fractal dimension $D_2$. (b)~Biorthogonal fractal dimension $D_2$. The green lines show states for the adjacency matrix without diagonal disorder ($W=0$); the red lines show states with diagonal disorder $W=1$.
    \redd{The solid lines represent numerical data for $N=1024$ and $64$ realizations, the dash-dotted lines corresponds to $N=4096$ and $16$ realizations.} The blue and purple dashed lines show states with ($W_{EP}=10^{-3}$) and without ($W_{EP}=0$) small diagonal disorder, for the case without $\epsilon$-perturbation of adjacency matrix (see~\ref{sec:nr:A} for details, the adjacency matrix is defective without $\epsilon$-perturbation and without diagonal disorder at studied parameters). (c)~Logarithmic scale of unnormalized spectral density with $\epsilon$-perturbation and $W=0$. The results of calculations for modified RRG $d=4$ (\mbox{$N=1024$} nodes, $64$ realizations) with reciprocity $r=0.1$ and hopping asymmetry $p=0.5$.}}  
\label{fig:RRG_ns}
\end{figure*}

\begin{figure*}[t]
\centering
    \includegraphics[width=1.0\textwidth]{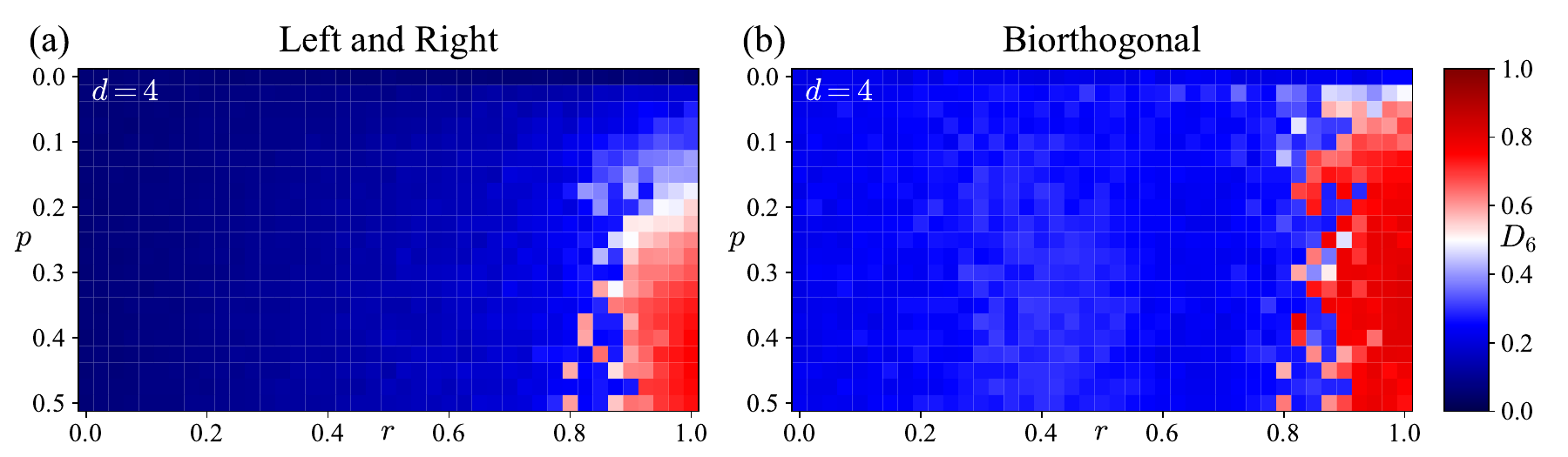}
    \caption{Numerical study of the single-particle Anderson transition for RRG using the fractal dimension $D_6$ as a localization measure calculated from average $IPR_q$ over the spectrum and all realizations. The results of calculations for networks with \mbox{$N=1024$} nodes are shown. Dependence of the left and right~(a) and biorthogonal~(b) fractal dimension $D_6$ on reciprocity $r$ and hopping asymmetry $p$ for $d=4$. Fractal dimensions are averaged over $16$ realizations, averaging over the left and right fractal dimensions calculated as independent.}  
\label{fig:RRG_D6}
\end{figure*}

\begin{figure*}[p]
    \centering
    \includegraphics[width=1\textwidth]{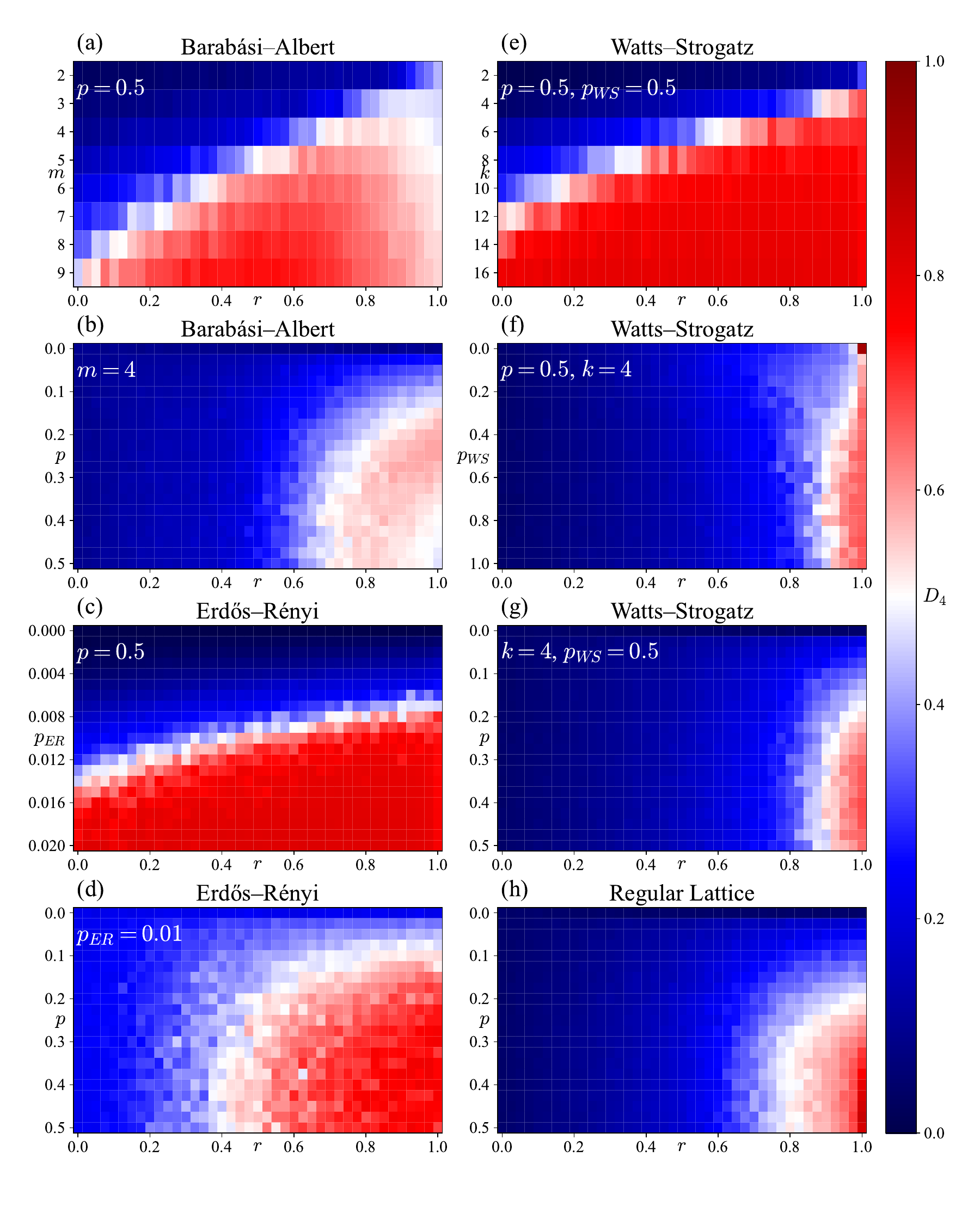}
    \caption{The emergence of localization in various models of random graphs ($N = 1024$) with varying hopping asymmetry $p$, reciprocity $r$ and specific parameters of models, considering left and right eigenvectors separately (see Section~\ref{sec:model} for a description of the specific parameters of each model). The fractal dimension $D_4$ is averaged for $8\times2$ samples per color cell. (a),(b)~BA model with different hopping asymmetry $p$, reciprocity $r$ and minimal node degree $m$. (c),(d)~ER model with different $p$, $r$ and the probability of the edge appearance $p_{ER}$. (e)-(g)~WS model with different $p$, $r$, number of connected neighbors in a ring topology $k$ for a node and the probability of the edge rewiring $p_{WS}$. (h)~RL with different $p$ and $r$.}
    \label{fig:D4models}
\end{figure*}

\subsection{Random graph models}
In numerical experiments we consider various models of random graphs as initial networks. We use the random regular graph (RRG) as the main model, it is a random $d$-regular graph without self-loops and parallel edges. The Erd\H{o}s-R\'enyi random graph (ER) is a $G_{n, p}$ model where each of the possible edges is chosen with probability $p_{ER}$~\cite{ER}. The Barab\'asi-Albert random graph (BA) is a model using Barab\'asi-Albert preferential attachment principle, where each new node has $m$ edges that are preferentially attached to existing nodes with higher degree~\cite{BA}. The Watts-Strogatz random graph (WS) is a model of network with small-world structure where each node is joined with its $k$ nearest neighbors in a ring topology and $p_{WS}$ is the probability of edge rewiring~\cite{WS}. The regular square lattice model (RL) is a two-dimensional grid graph, where each node is connected to its nearest neighbors with periodic boundary conditions. 

\section{Numerical Results}\label{sec:nr}
We study Anderson transition depending on graph reciprocity and hopping asymmetry on directed RRG, RL, ER, BA and WS networks. Whereas RRG is the standard model with a local tree-structure for MBL problem, BA, WS, ER are standard random graph models representing various features of real networks. We use the fractal dimension $D_4$ as a localization measure (see Section~\ref{sec:model} for details). 

\begin{figure*}[t]
    \centering
    \includegraphics[width=1\textwidth]{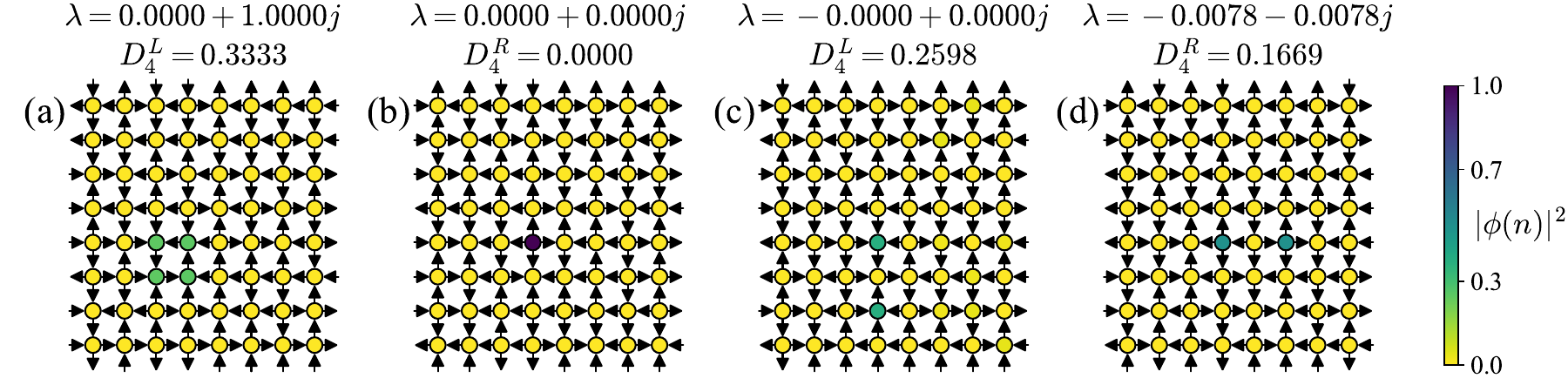}
    \caption{Examples of structural patterns that provide the emergence of localization transition in RL model with periodic boundary conditions ($8\times8$ nodes and $r=0$): (a) drain nodes, (b) source nodes, (c) TEN out, (d) TEN in. \red{Color plots show the square module of the eigenvector with normalization $\sum_n|\phi^{L(R)}(n)|^2=1$.}}
    \label{fig:sdio}
\end{figure*}

\subsection{Spectra and fractal dimensions}\label{sec:nr:A}

\subparagraph{RRG.}

Figure~\ref{fig:RRG_D4} shows the numerical simulations for the RRG model with $N=1024$. \red{To avoid the defectivity of the adjacency matrix due to the emergence of EP, we consider perturbation $\epsilon=10^{-5}$  around EP for numerical calculation (see the Section~\ref{sec:model} for a detailed description).} In the average fractal dimensions, the left and right fractal dimensions are included as independent, $D_q^{av}=\sum_j^{n^{real}}(D^L_{q,j}+D^R_{q,j})/(2n^{real})$, where $j$ is realization index and $n^{real}$ is the number of realizations. \red{This type of average is chosen to show a typical fractal dimension value for the system size, independent of the number of realizations.}

In RRG model with hoping asymmetry $p=0.5$ the Anderson's transition to the localized phase occurs when varying the reciprocity $r$ and nodes degree in both biorthogonal and separate cases (i.e., independent consideration of the left or right eigenvectors) (Fig.~\ref{fig:RRG_D4}(g),(h)). With an increase of node degree, the critical value of reciprocity decreases smoothly until the complete disappearance of localized states at values of degree $d \gtrsim 11$ for $N=1024$ for both biorthogonal and separate cases, which coincide with our analytical estimation (black line in Figure~\ref{fig:RRG_D4}(g)-(j)) calculated by (\ref{eq:r_c}). The transition also occurs with a violation of hopping asymmetry $p$, and localized states emerge in a fully bidirectional graph as the difference between $p$ and $1-p$ increases (Fig.~\ref{fig:RRG_D4}(i),(j)). There is also a region in parameter space where the emergence of localization is influenced \red{by both:} hopping asymmetry and reciprocity.

For separate eigenvectors, when $r<r_c$, localized states occur in the center of the adjacency matrix spectrum with the formation of a characteristic mobility edge (Fig.~\ref{fig:RRG_D4}(a),(c)-(f), see Fig.~\ref{fig:spec_rW_s} in Appendix~\ref{sec:A1}). For the biorthogonal case, the number of localized states is much smaller (Fig.~\ref{fig:RRG_D4}(b), see Fig.~\ref{fig:spec_rW_b} in Appendix~\ref{sec:A1}). The spectra in Figures~\ref{fig:RRG_D4}(a)-(c) are identical but colored by the left, biorthogonal, and right fractal dimensions, respectively. In the biorthogonal case, the left and right eigenvectors mutually suppress each other. As a consequence, the states are delocalized. \red{Similar effects happen in systems with non-Hermitian skin effects~\cite{HN_PhysRevB.58.8384,Bergholtz2021Exceptional,Okuma2023NonHermitian}}. For example, \red{in Hatano-Nelson chain}, where the left and right modes \red{are localized on the opposite edges of the chain}, their product is delocalized \red{in the bulk~\cite{HN_PhysRevB.58.8384}}. In our model, only a small number of states remain localized in the biorthogonal case. Several reasons for this are presented in Section~\ref{sec:ex}. Since the number of localized biorthogonal states is smaller than the number of separate ones and more nodes participate \red{in} biorthogonal localization (\red{nodes} where left and right eigenstates are localized), the biorthogonal fractal dimension is higher than the separate fractal dimension as presented in Figure~\ref{fig:RRG_D4}.

\red{Figure~\ref{fig:RRG_ns}(a),(b) demonstrates the distribution of fractal dimension at $r=0.1$ and $p=0.5$ for the $64$ realizations of RRG with $d=4$ and $N=1024$. A lot of left and right localized states are observed at EP (Figure~\ref{fig:RRG_ns}(a), purple line). The biorthogonal fractal dimension doesn't exist because the adjacency matrix is defective with the studied parameters. By adding $\epsilon$-perturbations (\redd{solid} green line) to prevent defectivity, the number of localized states significantly reduces. Previously isolated node structures with distinct states can now be connected, and eigenvectors spread across them. Consequently, the number of localized states decreases.} \redd{Note that the fractal dimension remains size-dependent for the studied number of nodes. For comparison, there are dash-dotted green lines that are calculated for the graphs with $N=4096$ nodes while maintaining other parameters. The occurrence of localization persists, but general $D_2(N)$ behavior requires further investigation.}

\red{Figure~\ref{fig:RRG_ns}(c) shows the number of states with unnormalized spectral density with $\epsilon$-perturbation. The majority of the states are located around localized states (graphs Figure~\ref{fig:RRG_D4}(b) and Figure~\ref{fig:RRG_ns}(c) have the same parameters).}

On Figure~\ref{fig:RRG_D6} the fractal dimension $D_6$ is calculated through average $IPR^{av}_\red{6}$ over spectrum and all realizations. \red{This type of averaging is chosen to show the sharpness of the transition. For fixed $r$ the same structural patterns of incoming and outgoing edges are forming independently of $p$ value. In this averaging we have the range of states with a total number of $N\times n^{real}$ for each $p$. For the fixed $r$ only changing the hopping asymmetry influences the average $IPR_6$. Thus, Fig.~\ref{fig:RRG_D6} shows how $p$ effects the localization properties.}

The transition in the reciprocity parameter $r$ is sharp, while when $p$ changes, the estimated value of the fractal dimension changes smoothly for both biorthogonal and separate cases. The reasons are related to the difference between the mechanisms by which parameters influence the distribution of the wave function. Reciprocity changes the probability of occurrence of certain connectivity partners in the network, while the localization emergence follows from the presence or absence of a certain connection, and as a result the transition zone is sharp. \red{With a change in the value of $p$} the fractal dimension continuously changes from fully localized state at EP $p=0$ to fully delocalized. Consequently, the hopping asymmetry $p$ affects the length of the localization. The eigenvector of a localized state takes finite non-zero values not only on the in or out nodes, like at EP. Therefore, left and right eigenvectors can suppress each other, and biorthogonal localization is more sensitive to changes in $p$, which leads to an increase in biorthogonal fractal dimension.

\subparagraph{Other models.} 

The Anderson transition under the reciprocity parameter $r$ and the hopping asymmetry $p$ is also observed for a regular lattice with periodic boundary conditions, ER, BA and WS models (Fig.~\ref{fig:D4models}). For a scale-free BA network localized states are not observed at large values of the average degree of a vertex in the model. The phenomenon of percolation is described in detail for the ER model, and the percolation threshold can be calculated analytically as a function of network size and density \red{\cite{Palla2007}}. Localized states in the ER model are observed for all values of the reciprocity parameter above the percolation threshold for cliques of order 2 (edges), $p_{ER}^{percol}=1/N\approx0.001$, and are limited from above by some critical network density, the value of which depends slightly on reciprocity. In the Barab\'asi-Albert model, localized states can also be observed at heavy nodes in the case of a completely undirected graph which is described in detail in \cite{PastorSatorras2015Distinct}.

\subsection{Origin of localization}\label{sec:ool}

\red{Here} we show structural patterns that provide the occurrence of localization transition. 2D regular grid with periodic boundary conditions with $8\times8$ nodes and $r=0$ is examined as an example. In our research we have found two types of localization:
\begin{itemize}
    \item Drain and source nodes for left and right eigenvectors (Fig.~\ref{fig:sdio}(a),(b));
    \item Out and in topological equivalent nodes (TEN) for left and right eigenvectors (Fig.~\ref{fig:sdio}(c),(d)).
\end{itemize}

\subparagraph{Drain and source nodes.} 
A set of nodes connected to an external graph with edges pointing in the same direction (only in or out) can be the origin of localization. We call those structures drain (source) nodes for domination of incoming (outgoing) edges. These nodes have an analogy in the context of non-Hermitian skin effect; they correspond to the leftmost or rightmost node in Hatano-Nelson or the non-Hermitian Su-Schrieffer-Heeger chains with an open boundary condition \cite{Kunst2018Biorthogonal,Yao2018Edge,Bergholtz2021Exceptional}. The appearance of the localization depends on hopping asymmetry (Fig.~\ref{fig:RRG_D4}) and the number of nodes in the drain (source) set $n$. For unidirectional edges, if $n$ is small compared to the total number of nodes in the graph, the eigenvector is \red{distributed} only on the set nodes. The fractal dimension of the left (right) state is $D_q^{L(R)}\sim\log_2(n)/\log_2(N)$ for drain (source). If the graph has drain and source nodes with the same eigenvalue, then the product of the left and right eigenvectors can be localized too, but it needs to have a higher hopping asymmetry due to mutual suppression (see Sec.~\ref{sec:ex}). At the EP, the simultaneous existence of the drain and source structures makes the graph adjacency matrix defective. This issue is discussed in detail in Section~\ref{sec:ex}.

\subparagraph{TENs.} 
If several nodes have all out- or in-edges connected to the same set of other nodes, then the localization on these nodes appears in the left or right eigenvector, respectively. Such nodes are called topologically equivalent (TEN). TEN for undirected RRG was found in \cite{Kochergin2023Anatomy}. On an undirected graph, where in and out neighbors are identical, TEN's eigenfunction is non-zero only on TEN nodes, and the eigenvalues of the unconnected to each other TEN nodes are exactly equal to zero. On a directed graph, TEN states can form a band around zero in the complex plane (see Fig.~\ref{fig:sdio}\red{(c),}(d)). In the case of an undirected graph, the equation for each vertex is $\sum_j \psi^{NN}_j=\lambda \psi_i$, where $i$ runs over TEN nodes and $\sum_j\psi^{NN}_j$ is the sum over nearest neighbors nodes. In directed case the condition is not so strict. The most probable TEN constructions \red{are} shown in the Figure~\ref{fig:sdio}\red{(c),}(d). Except for simple TENs consisting of two nodes, more complex TENs are possible. They don't necessarily have to be TEN pairwise, but each of our neighbors has at least two neighbors from a TEN cluster. 

There is a possibility of biorthogonal localization to be caused by TENs as well. In this case vertices included in TENs must have both incoming and outgoing common neighbors, which makes this case similar to an undirected graph. The eigenvalue for this state will be real. The formation of such a structure is impossible on a lattice. In comparison to the first case, such structures do not require any closeness to EP (i.e. they do not require adding feedback correction $\epsilon$ to make an adjacency matrix diagonalizable). Also, from the point of view of random walks on the graph, paths through such vertices are not dead ends.

The determination of localization structures on the other random graph models is far more difficult due to the complex nature of the network. However, we suppose that the eigenfunction will be localized on the nodes with the properties presented above or their combinations.

\begin{figure*}[t]
    \centering
    \includegraphics[width=1\textwidth]{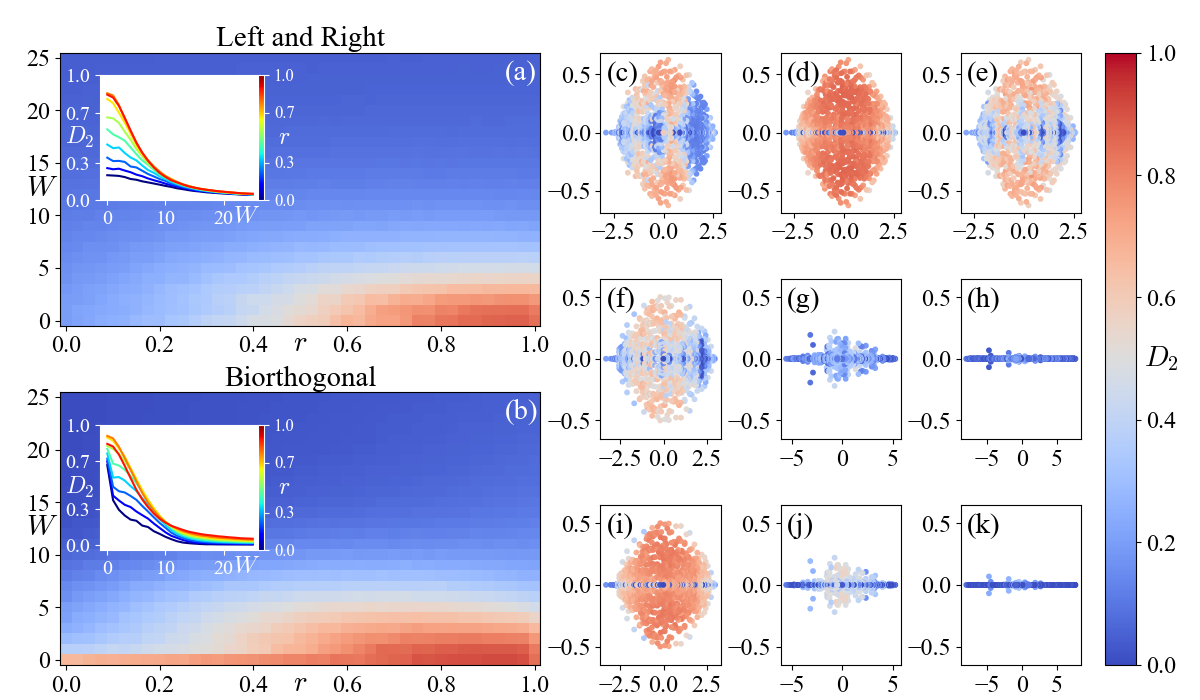}
    \caption{(a)~Left and right fractal dimensions $D_2$ for RRG for different $r$ and $W$, $p=0.5$; cuts by $r$ (inset). (b)~Biorthogonal fractal dimensions $D_2$ for RRG for different $r$ and $W$, $p=0.5$; cuts by $r$ (inset).
    Spectra are colored by fractal dimension $D_2$: left~(c), biorthogonal~(d), and right~(e) fractal dimensions for the same realization, $W=4$, $r=0.375$; right~(f),~(g),~(h) and biorthogonal~(i),~(j),~(k) fractal dimensions with different absolute values of diagonal disorder, $W=5,10,15$, $r=0.25$. \red{The system size is $N=1024$.}}
    \label{fig:rW}
\end{figure*}

\section{Diagonal disorder}\label{sec:dd}

In this Section we consider diagonal disorder on $rp$-network presented in Section 2. Since RRG with diagonal disorder is a toy model for the many-body localization problem, we use it as an initial undirected graph.

We consider spinless fermion with diagonal disorder on a directed graph described by Hamiltonian with matrix element\red{~\cite{anderson1958,abou1973selfconsistent,DeTomasi2022NonHermitian,Luo2021Universality,Luo2022Unifying}}:
\begin{equation}\label{eq:H_W}
    H_{ij}=t_{ij}A_{ij}+\varepsilon_i \delta_{ij}
\end{equation}
where $\varepsilon_i$ are uniformly distributed on $[-W/2,W/2]$. \red{Commonly, RRG with vertex degree $d=3$ is used, with corresponding critical value of diagonal disorder $W_{c}\approx 18.17$~\cite{Luca2014,Kravtsov2018nonergodic,Parisi2019anderson,Tikhonov2021From}}. In our case with hopping symmetry $p=0.5$, \red{at $r=1$} the value of the criticality limit should decrease by half, \red{because all edges are undirected and have a weight of $0.5$}.  
\red{In our $rp$-network based on RRG with $d=3$, the directed graph without diagonal disorder has a lot of degenerate states at $r=0$, because a lot of elementary one-node source and drain configurations appear. Furthermore, localized states are observed in almost all range of reciprocity values, as it goes from Eq.~(\ref{eq:r_c}) and Fig.~\ref{fig:RRG_D4}(g),(h)}. To decrease the number \red{of degenerate states} $d=4$ is \red{investigated}. \red{The system size is $N=1024$.}

We studied the Anderson transition for both separate and biorthogonal states while simultaneously varying network reciprocity and diagonal disorder amplitude in RRG (Fig.~\ref{fig:rW}, Fig.~\ref{fig:spec_rW_s} and Fig.~\ref{fig:spec_rW_b}).

\red{From numerical results (Fig.~\ref{fig:rW}(b)) we see} the sharp transition of the biorthogonal fractal dimension for $r<0.5$ \red{($N=1024$)} is caused by presence of a weak diagonal disorder.
\red{At EP, weak random diagonal disorder eliminates the defectiveness of the non-Hermitian matrix. With the diagonal disorder, each localized structural pattern has its own eigenstate with different eigenvalues.
The blue dashed line in Figure~\ref{fig:RRG_ns} demonstrates fractal dimension $D_2$ at $W_{EP}\sim10^{-3}$ without $\epsilon$-perturbation of the adjacency matrix.}

\red{Around EP, in the presence of weak non-zero disorder, the number of biorthogonal localized states increases compared to the disorder-free situation (the red and green lines in Fig.~\ref{fig:RRG_ns}(b)).
Considering the first-order perturbation theory of Hamiltonian~(\ref{eq:H_W}), where the second term is unperturbed Hamiltonian (diagonal on-site disorder) and the first term is perturbation (adjacency matrix), the zeroth order eigenvectors localized on-site, $\psi^{0}_i(j)=\delta_{ji}$. The first order perturbation eigenvector is
\begin{equation}
    \psi^{1}_i(j)=\delta_{ji}+\frac{A_{ji}}{\varepsilon_i-\varepsilon_j}.
\end{equation}
One of the localization criteria is the presence of only a small number of isolated resonances (resonances at $A_{ji}>\varepsilon_i-\varepsilon_j$ (see Section~3.2 in~\cite{scardicchio2017perturbation})). If there are many resonances and they overlap, the eigenvector is distributed on many nodes. Since some of our edges have weight $A_{kl}=\epsilon=10^{-5}$, small perturbations begin to break resonances and the number of the localized states increases. }

Independently of the fractal dimension type, when the amplitude of the diagonal disorder increases above the critical value ($W_{c} \sim 10$ for RRG), all states become localized independently of reciprocity, but they may still retain complex eigenvalues \red{(Fig.~\ref{fig:rW}(g),(j))}. With further increase in diagonal disorder amplitude $W$, the spectrum is squeezed to the real axis (Fig.~\ref{fig:rW}(h),(k)) and all states become real-valued, which is observed around $W_c \sim 20$ (Fig.~\ref{fig:rW}(a),(b)). 

When the value of the disorder amplitude is around $W\sim5$ and reciprocity around $r\sim0.25$, the mutual influence of network reciprocity and disorder can be observed. For the separate left and right eigenvectors patterns of alternation of localized and non-localized states in the distribution of the phases over the spectrum along the real axis are found (Fig.~\ref{fig:rW}(c),(e),(f), Fig.~\ref{fig:spec_rW_s}). This phenomenon may be a consequence of the mutual influence of diagonal disorder and topological structures born from a small value of reciprocity. The phenomenon of interchange of delocalized and localized states with several mobility edges in intermediate diagonal disorder was found in the Hermitian system both experimentally~\cite{gao2023experimental} and theoretically~\cite{goncalves2023quasiperiodicity}.

\section{Conclusion}\label{sec:con}

In the present study, we investigated the single-particle Anderson localization problem for non-Hermitian systems considering a smooth transition from an undirected to a fully directed graph with varying reciprocity parameter $r$ and hopping asymmetry $p$. We observed the emergence of localized states with an increase in the proportion of unidirectional edges and an increase in hopping asymmetry near EP both on left and right eigenvectors separately, as well as on their biorthogonal product. For separate eigenvectors, the eigenvalues corresponding to localized states cluster near the center of the spectrum with the formation of a characteristic mobility edge. Biorthogonal localized states affect the transport properties of the system since the Green's function contains both left and right eigenvectors.

Additionally, the combination of structural disorder caused by graph reciprocity and diagonal disorder is calculated numerically. It turned out that left and right fractal dimensions have a region around $r\sim0.25$, $W\sim5$ \red{for $N=1024$} with complex interchange of spectral stripes of localized and delocalized states. Similar effects exist in the Hermitian case \cite{gao2023experimental,goncalves2023quasiperiodicity}, but for non-Hermitian systems they were found for the first time. For the biorthogonal case, a significant change of the fractal dimension with the addition of small diagonal disorder was shown (\red{Fig.~\ref{fig:RRG_ns} and} Fig.~\ref{fig:rW}(d)).

The presence of localization in the vicinity of EP is similar to that of non-Hermitian skin effect. However, the latter is also characterized by topological invariants~\cite{Bergholtz2021Exceptional}. The question of whether they exist and what they are in the case of random graphs requires further investigation. Another problem that could be solved in the future is analytical study of the spectral density for the $rp$-networks. Since it is known that spectra of non-hermitian systems with open, semi-infinite, and periodic boundary conditions are different~\cite{Okuma2023NonHermitian} and our model has an analog of a boundary, the spectral density can differ when compared to the graph without source or drain nodes that were found in~\cite{Neri2012Spectra}.

Another interesting problem is the criticality indexes of Anderson transition of the $rp$-networks. From the topological point of view, if the graph has isolated regions and doesn't have strong connectivity, it leads to the question of cluster percolation dependence on reciprocity. Moreover, it will be necessary to determine the correspondence between directed graphs and open MBL systems.

\subparagraph{Acknowledgements.}
We are grateful to Tagir Aushev for the inspiring question and to Alexander Gorsky and Ivan M. Khaymovich for the valuable discussion.

\bibliography{bib_PRE_resub.bib}

\begin{thebibliography}{73}%
\makeatletter
\providecommand \@ifxundefined [1]{%
 \@ifx{#1\undefined}
}%
\providecommand \@ifnum [1]{%
 \ifnum #1\expandafter \@firstoftwo
 \else \expandafter \@secondoftwo
 \fi
}%
\providecommand \@ifx [1]{%
 \ifx #1\expandafter \@firstoftwo
 \else \expandafter \@secondoftwo
 \fi
}%
\providecommand \natexlab [1]{#1}%
\providecommand \enquote  [1]{``#1''}%
\providecommand \bibnamefont  [1]{#1}%
\providecommand \bibfnamefont [1]{#1}%
\providecommand \citenamefont [1]{#1}%
\providecommand \href@noop [0]{\@secondoftwo}%
\providecommand \href [0]{\begingroup \@sanitize@url \@href}%
\providecommand \@href[1]{\@@startlink{#1}\@@href}%
\providecommand \@@href[1]{\endgroup#1\@@endlink}%
\providecommand \@sanitize@url [0]{\catcode `\\12\catcode `\$12\catcode
  `\&12\catcode `\#12\catcode `\^12\catcode `\_12\catcode `\%12\relax}%
\providecommand \@@startlink[1]{}%
\providecommand \@@endlink[0]{}%
\providecommand \url  [0]{\begingroup\@sanitize@url \@url }%
\providecommand \@url [1]{\endgroup\@href {#1}{\urlprefix }}%
\providecommand \urlprefix  [0]{URL }%
\providecommand \Eprint [0]{\href }%
\providecommand \doibase [0]{https://doi.org/}%
\providecommand \selectlanguage [0]{\@gobble}%
\providecommand \bibinfo  [0]{\@secondoftwo}%
\providecommand \bibfield  [0]{\@secondoftwo}%
\providecommand \translation [1]{[#1]}%
\providecommand \BibitemOpen [0]{}%
\providecommand \bibitemStop [0]{}%
\providecommand \bibitemNoStop [0]{.\EOS\space}%
\providecommand \EOS [0]{\spacefactor3000\relax}%
\providecommand \BibitemShut  [1]{\csname bibitem#1\endcsname}%
\let\auto@bib@innerbib\@empty
\bibitem [{\citenamefont {Altshuler}\ \emph {et~al.}(1997)\citenamefont
  {Altshuler}, \citenamefont {Gefen}, \citenamefont {Kamenev},\ and\
  \citenamefont {Levitov}}]{altshuler1997quasiparticle}%
  \BibitemOpen
  \bibfield  {author} {\bibinfo {author} {\bibfnamefont {B.~L.}\ \bibnamefont
  {Altshuler}}, \bibinfo {author} {\bibfnamefont {Y.}~\bibnamefont {Gefen}},
  \bibinfo {author} {\bibfnamefont {A.}~\bibnamefont {Kamenev}},\ and\ \bibinfo
  {author} {\bibfnamefont {L.~S.}\ \bibnamefont {Levitov}},\ }\bibfield
  {title} {\bibinfo {title} {Quasiparticle lifetime in a finite system: A
  nonperturbative approach},\ }\href@noop {} {\bibfield  {journal} {\bibinfo
  {journal} {Physical review letters}\ }\textbf {\bibinfo {volume} {78}},\
  \bibinfo {pages} {2803} (\bibinfo {year} {1997})}\BibitemShut {NoStop}%
\bibitem [{\citenamefont {Abou-Chacra}\ \emph {et~al.}(1973)\citenamefont
  {Abou-Chacra}, \citenamefont {Thouless},\ and\ \citenamefont
  {Anderson}}]{abou1973selfconsistent}%
  \BibitemOpen
  \bibfield  {author} {\bibinfo {author} {\bibfnamefont {R.}~\bibnamefont
  {Abou-Chacra}}, \bibinfo {author} {\bibfnamefont {D.~J.}\ \bibnamefont
  {Thouless}},\ and\ \bibinfo {author} {\bibfnamefont {P.~W.}\ \bibnamefont
  {Anderson}},\ }\bibfield  {title} {\bibinfo {title} {A selfconsistent theory
  of localization},\ }\href@noop {} {\bibfield  {journal} {\bibinfo  {journal}
  {Journal of Physics C: Solid State Physics}\ }\textbf {\bibinfo {volume}
  {6}},\ \bibinfo {pages} {1734} (\bibinfo {year} {1973})}\BibitemShut
  {NoStop}%
\bibitem [{\citenamefont {Tikhonov}\ and\ \citenamefont
  {Mirlin}(2021)}]{Tikhonov2021From}%
  \BibitemOpen
  \bibfield  {author} {\bibinfo {author} {\bibfnamefont {K.~S.}\ \bibnamefont
  {Tikhonov}}\ and\ \bibinfo {author} {\bibfnamefont {A.~D.}\ \bibnamefont
  {Mirlin}},\ }\bibfield  {title} {\bibinfo {title} {From anderson localization
  on random regular graphs to many-body localization},\ }\href
  {https://doi.org/https://doi.org/10.1016/j.aop.2021.168525} {\bibfield
  {journal} {\bibinfo  {journal} {Annals of Physics}\ }\textbf {\bibinfo
  {volume} {435}},\ \bibinfo {pages} {168525} (\bibinfo {year} {2021})},\
  \bibinfo {note} {special Issue on Localisation 2020}\BibitemShut {NoStop}%
\bibitem [{\citenamefont {Wegner}(1979)}]{Wegner1979Disordered}%
  \BibitemOpen
  \bibfield  {author} {\bibinfo {author} {\bibfnamefont {F.~J.}\ \bibnamefont
  {Wegner}},\ }\bibfield  {title} {\bibinfo {title} {Disordered system with $n$
  orbitals per site: $n=\ensuremath{\infty}$ limit},\ }\href
  {https://doi.org/10.1103/PhysRevB.19.783} {\bibfield  {journal} {\bibinfo
  {journal} {Phys. Rev. B}\ }\textbf {\bibinfo {volume} {19}},\ \bibinfo
  {pages} {783} (\bibinfo {year} {1979})}\BibitemShut {NoStop}%
\bibitem [{\citenamefont {Smith}\ \emph {et~al.}(2017)\citenamefont {Smith},
  \citenamefont {Knolle}, \citenamefont {Kovrizhin},\ and\ \citenamefont
  {Moessner}}]{Smith2017Disorder}%
  \BibitemOpen
  \bibfield  {author} {\bibinfo {author} {\bibfnamefont {A.}~\bibnamefont
  {Smith}}, \bibinfo {author} {\bibfnamefont {J.}~\bibnamefont {Knolle}},
  \bibinfo {author} {\bibfnamefont {D.~L.}\ \bibnamefont {Kovrizhin}},\ and\
  \bibinfo {author} {\bibfnamefont {R.}~\bibnamefont {Moessner}},\ }\bibfield
  {title} {\bibinfo {title} {Disorder-free localization},\ }\href
  {https://doi.org/10.1103/PhysRevLett.118.266601} {\bibfield  {journal}
  {\bibinfo  {journal} {Phys. Rev. Lett.}\ }\textbf {\bibinfo {volume} {118}},\
  \bibinfo {pages} {266601} (\bibinfo {year} {2017})}\BibitemShut {NoStop}%
\bibitem [{\citenamefont {Kutlin}\ and\ \citenamefont
  {Khaymovich}(2020)}]{Kutlin20220Renormalization}%
  \BibitemOpen
  \bibfield  {author} {\bibinfo {author} {\bibfnamefont {A.~G.}\ \bibnamefont
  {Kutlin}}\ and\ \bibinfo {author} {\bibfnamefont {I.~M.}\ \bibnamefont
  {Khaymovich}},\ }\bibfield  {title} {\bibinfo {title} {{Renormalization to
  localization without a small parameter}},\ }\href
  {https://doi.org/10.21468/SciPostPhys.8.4.049} {\bibfield  {journal}
  {\bibinfo  {journal} {SciPost Phys.}\ }\textbf {\bibinfo {volume} {8}},\
  \bibinfo {pages} {049} (\bibinfo {year} {2020})}\BibitemShut {NoStop}%
\bibitem [{\citenamefont {Goetschy}\ and\ \citenamefont
  {Skipetrov}(2013)}]{goetschy2013euclidean}%
  \BibitemOpen
  \bibfield  {author} {\bibinfo {author} {\bibfnamefont {A.}~\bibnamefont
  {Goetschy}}\ and\ \bibinfo {author} {\bibfnamefont {S.~E.}\ \bibnamefont
  {Skipetrov}},\ }\href@noop {} {\bibinfo {title} {Euclidean random matrices
  and their applications in physics}} (\bibinfo {year} {2013}),\ \Eprint
  {https://arxiv.org/abs/arXiv:1303.2880} {arXiv:arXiv:1303.2880 [math-ph]}
  \BibitemShut {NoStop}%
\bibitem [{\citenamefont {Biroli}\ and\ \citenamefont
  {Monasson}(1999)}]{Biroli1999Asingle}%
  \BibitemOpen
  \bibfield  {author} {\bibinfo {author} {\bibfnamefont {G.}~\bibnamefont
  {Biroli}}\ and\ \bibinfo {author} {\bibfnamefont {R.}~\bibnamefont
  {Monasson}},\ }\bibfield  {title} {\bibinfo {title} {A single defect
  approximation for localized states on random lattices},\ }\href
  {https://doi.org/10.1088/0305-4470/32/24/101} {\bibfield  {journal} {\bibinfo
   {journal} {Journal of Physics A: Mathematical and General}\ }\textbf
  {\bibinfo {volume} {32}},\ \bibinfo {pages} {L255} (\bibinfo {year}
  {1999})}\BibitemShut {NoStop}%
\bibitem [{\citenamefont {Pastor-Satorras}\ and\ \citenamefont
  {Castellano}(2015)}]{PastorSatorras2015Distinct}%
  \BibitemOpen
  \bibfield  {author} {\bibinfo {author} {\bibfnamefont {R.}~\bibnamefont
  {Pastor-Satorras}}\ and\ \bibinfo {author} {\bibfnamefont {C.}~\bibnamefont
  {Castellano}},\ }\bibfield  {title} {\bibinfo {title} {Distinct types of
  eigenvector localization in networks},\ }\href
  {https://doi.org/10.1038/srep18847} {\bibfield  {journal} {\bibinfo
  {journal} {Scientific Reports}\ }\textbf {\bibinfo {volume} {6}} (\bibinfo
  {year} {2015})}\BibitemShut {NoStop}%
\bibitem [{\citenamefont {Nechaev}\ \emph {et~al.}(2017)\citenamefont
  {Nechaev}, \citenamefont {Tamm},\ and\ \citenamefont
  {Valba}}]{Nechaev2017Path}%
  \BibitemOpen
  \bibfield  {author} {\bibinfo {author} {\bibfnamefont {S.~K.}\ \bibnamefont
  {Nechaev}}, \bibinfo {author} {\bibfnamefont {M.~V.}\ \bibnamefont {Tamm}},\
  and\ \bibinfo {author} {\bibfnamefont {O.~V.}\ \bibnamefont {Valba}},\
  }\bibfield  {title} {\bibinfo {title} {Path counting on simple graphs: from
  escape to localization},\ }\href {https://doi.org/10.1088/1742-5468/aa680a}
  {\bibfield  {journal} {\bibinfo  {journal} {Journal of Statistical Mechanics:
  Theory and Experiment}\ }\textbf {\bibinfo {volume} {2017}},\ \bibinfo
  {pages} {053301} (\bibinfo {year} {2017})}\BibitemShut {NoStop}%
\bibitem [{\citenamefont {Matyushina}(2023)}]{matyushina2023statistics}%
  \BibitemOpen
  \bibfield  {author} {\bibinfo {author} {\bibfnamefont {Z.~D.}\ \bibnamefont
  {Matyushina}},\ }\href@noop {} {\bibinfo {title} {Statistics of paths on
  graphs with two heavy roots}} (\bibinfo {year} {2023}),\ \Eprint
  {https://arxiv.org/abs/arXiv:2302.05876} {arXiv:arXiv:2302.05876
  [cond-mat.stat-mech]} \BibitemShut {NoStop}%
\bibitem [{\citenamefont {Avetisov}\ \emph {et~al.}(2019)\citenamefont
  {Avetisov}, \citenamefont {Gorsky}, \citenamefont {Nechaev},\ and\
  \citenamefont {Valba}}]{Avetisov2019Localization}%
  \BibitemOpen
  \bibfield  {author} {\bibinfo {author} {\bibfnamefont {V.}~\bibnamefont
  {Avetisov}}, \bibinfo {author} {\bibfnamefont {A.}~\bibnamefont {Gorsky}},
  \bibinfo {author} {\bibfnamefont {S.}~\bibnamefont {Nechaev}},\ and\ \bibinfo
  {author} {\bibfnamefont {O.}~\bibnamefont {Valba}},\ }\bibfield  {title}
  {\bibinfo {title} {{Localization and non-ergodicity in clustered random
  networks}},\ }\href {https://doi.org/10.1093/comnet/cnz026} {\bibfield
  {journal} {\bibinfo  {journal} {Journal of Complex Networks}\ }\textbf
  {\bibinfo {volume} {8}},\ \bibinfo {pages} {cnz026} (\bibinfo {year}
  {2019})},\ \Eprint
  {https://arxiv.org/abs/https://academic.oup.com/comnet/article-pdf/8/2/cnz026/33543561/cnz026.pdf}
  {https://academic.oup.com/comnet/article-pdf/8/2/cnz026/33543561/cnz026.pdf}
  \BibitemShut {NoStop}%
\bibitem [{\citenamefont {Valba}\ and\ \citenamefont
  {Gorsky}(2021)}]{Valba2021Interacting}%
  \BibitemOpen
  \bibfield  {author} {\bibinfo {author} {\bibfnamefont {O.}~\bibnamefont
  {Valba}}\ and\ \bibinfo {author} {\bibfnamefont {A.}~\bibnamefont {Gorsky}},\
  }\bibfield  {title} {\bibinfo {title} {Interacting thermofield doubles and
  critical behavior in random regular graphs},\ }\href
  {https://doi.org/10.1103/PhysRevD.103.106013} {\bibfield  {journal} {\bibinfo
   {journal} {Phys. Rev. D}\ }\textbf {\bibinfo {volume} {103}},\ \bibinfo
  {pages} {106013} (\bibinfo {year} {2021})}\BibitemShut {NoStop}%
\bibitem [{\citenamefont {Kochergin}\ \emph
  {et~al.}(2023{\natexlab{a}})\citenamefont {Kochergin}, \citenamefont
  {Khaymovich}, \citenamefont {Valba},\ and\ \citenamefont
  {Gorsky}}]{Kochergin2023Anatomy}%
  \BibitemOpen
  \bibfield  {author} {\bibinfo {author} {\bibfnamefont {D.}~\bibnamefont
  {Kochergin}}, \bibinfo {author} {\bibfnamefont {I.~M.}\ \bibnamefont
  {Khaymovich}}, \bibinfo {author} {\bibfnamefont {O.}~\bibnamefont {Valba}},\
  and\ \bibinfo {author} {\bibfnamefont {A.}~\bibnamefont {Gorsky}},\
  }\bibfield  {title} {\bibinfo {title} {Anatomy of the fragmented hilbert
  space: Eigenvalue tunneling, quantum scars, and localization in the perturbed
  random regular graph},\ }\href {https://doi.org/10.1103/PhysRevB.108.094203}
  {\bibfield  {journal} {\bibinfo  {journal} {Phys. Rev. B}\ }\textbf {\bibinfo
  {volume} {108}},\ \bibinfo {pages} {094203} (\bibinfo {year}
  {2023}{\natexlab{a}})}\BibitemShut {NoStop}%
\bibitem [{\citenamefont {Valba}\ and\ \citenamefont
  {Gorsky}(2022)}]{Valba2022Mobility}%
  \BibitemOpen
  \bibfield  {author} {\bibinfo {author} {\bibfnamefont {O.}~\bibnamefont
  {Valba}}\ and\ \bibinfo {author} {\bibfnamefont {A.}~\bibnamefont {Gorsky}},\
  }\bibfield  {title} {\bibinfo {title} {{Mobility edge in the Anderson model
  on partially disordered random regular graphs}},\ }\href
  {https://doi.org/10.31857/S1234567822180094} {\bibfield  {journal} {\bibinfo
  {journal} {Pisma Zh. Eksp. Teor. Fiz.}\ }\textbf {\bibinfo {volume} {116}},\
  \bibinfo {pages} {392} (\bibinfo {year} {2022})},\ \Eprint
  {https://arxiv.org/abs/2112.14585} {arXiv:2112.14585 [cond-mat.dis-nn]}
  \BibitemShut {NoStop}%
\bibitem [{\citenamefont {Kochergin}\ \emph
  {et~al.}(2023{\natexlab{b}})\citenamefont {Kochergin}, \citenamefont
  {Khaymovich}, \citenamefont {Valba},\ and\ \citenamefont
  {Gorsky}}]{Kochergin2023Robust}%
  \BibitemOpen
  \bibfield  {author} {\bibinfo {author} {\bibfnamefont {D.}~\bibnamefont
  {Kochergin}}, \bibinfo {author} {\bibfnamefont {I.~M.}\ \bibnamefont
  {Khaymovich}}, \bibinfo {author} {\bibfnamefont {O.}~\bibnamefont {Valba}},\
  and\ \bibinfo {author} {\bibfnamefont {A.}~\bibnamefont {Gorsky}},\
  }\href@noop {} {\bibinfo {title} {Robust extended states in anderson model on
  partially disordered random regular graphs}} (\bibinfo {year}
  {2023}{\natexlab{b}}),\ \Eprint {https://arxiv.org/abs/arXiv:2309.05691}
  {arXiv:arXiv:2309.05691 [cond-mat.dis-nn]} \BibitemShut {NoStop}%
\bibitem [{\citenamefont {Brunel}(2000)}]{brunel2000dynamics}%
  \BibitemOpen
  \bibfield  {author} {\bibinfo {author} {\bibfnamefont {N.}~\bibnamefont
  {Brunel}},\ }\bibfield  {title} {\bibinfo {title} {Dynamics of sparsely
  connected networks of excitatory and inhibitory spiking neurons},\
  }\href@noop {} {\bibfield  {journal} {\bibinfo  {journal} {Journal of
  computational neuroscience}\ }\textbf {\bibinfo {volume} {8}},\ \bibinfo
  {pages} {183} (\bibinfo {year} {2000})}\BibitemShut {NoStop}%
\bibitem [{\citenamefont {Bascompte}(2009)}]{bascompte2009disentangling}%
  \BibitemOpen
  \bibfield  {author} {\bibinfo {author} {\bibfnamefont {J.}~\bibnamefont
  {Bascompte}},\ }\bibfield  {title} {\bibinfo {title} {Disentangling the web
  of life},\ }\href@noop {} {\bibfield  {journal} {\bibinfo  {journal}
  {Science}\ }\textbf {\bibinfo {volume} {325}},\ \bibinfo {pages} {416}
  (\bibinfo {year} {2009})}\BibitemShut {NoStop}%
\bibitem [{\citenamefont {Milo}\ \emph {et~al.}(2002)\citenamefont {Milo},
  \citenamefont {Shen-Orr}, \citenamefont {Itzkovitz}, \citenamefont {Kashtan},
  \citenamefont {Chklovskii},\ and\ \citenamefont {Alon}}]{milo2002network}%
  \BibitemOpen
  \bibfield  {author} {\bibinfo {author} {\bibfnamefont {R.}~\bibnamefont
  {Milo}}, \bibinfo {author} {\bibfnamefont {S.}~\bibnamefont {Shen-Orr}},
  \bibinfo {author} {\bibfnamefont {S.}~\bibnamefont {Itzkovitz}}, \bibinfo
  {author} {\bibfnamefont {N.}~\bibnamefont {Kashtan}}, \bibinfo {author}
  {\bibfnamefont {D.}~\bibnamefont {Chklovskii}},\ and\ \bibinfo {author}
  {\bibfnamefont {U.}~\bibnamefont {Alon}},\ }\bibfield  {title} {\bibinfo
  {title} {Network motifs: simple building blocks of complex networks},\
  }\href@noop {} {\bibfield  {journal} {\bibinfo  {journal} {Science}\ }\textbf
  {\bibinfo {volume} {298}},\ \bibinfo {pages} {824} (\bibinfo {year}
  {2002})}\BibitemShut {NoStop}%
\bibitem [{\citenamefont {Kwak}\ \emph {et~al.}(2010)\citenamefont {Kwak},
  \citenamefont {Lee}, \citenamefont {Park},\ and\ \citenamefont
  {Moon}}]{kwak2010twitter}%
  \BibitemOpen
  \bibfield  {author} {\bibinfo {author} {\bibfnamefont {H.}~\bibnamefont
  {Kwak}}, \bibinfo {author} {\bibfnamefont {C.}~\bibnamefont {Lee}}, \bibinfo
  {author} {\bibfnamefont {H.}~\bibnamefont {Park}},\ and\ \bibinfo {author}
  {\bibfnamefont {S.}~\bibnamefont {Moon}},\ }\bibfield  {title} {\bibinfo
  {title} {What is twitter, a social network or a news media?},\ }in\
  \href@noop {} {\emph {\bibinfo {booktitle} {Proceedings of the 19th
  international conference on World wide web}}}\ (\bibinfo {year} {2010})\ pp.\
  \bibinfo {pages} {591--600}\BibitemShut {NoStop}%
\bibitem [{\citenamefont {Broder}\ \emph {et~al.}(2000)\citenamefont {Broder},
  \citenamefont {Kumar}, \citenamefont {Maghoul}, \citenamefont {Raghavan},
  \citenamefont {Rajagopalan}, \citenamefont {Stata}, \citenamefont {Tomkins},\
  and\ \citenamefont {Wiener}}]{broder2000graph}%
  \BibitemOpen
  \bibfield  {author} {\bibinfo {author} {\bibfnamefont {A.}~\bibnamefont
  {Broder}}, \bibinfo {author} {\bibfnamefont {R.}~\bibnamefont {Kumar}},
  \bibinfo {author} {\bibfnamefont {F.}~\bibnamefont {Maghoul}}, \bibinfo
  {author} {\bibfnamefont {P.}~\bibnamefont {Raghavan}}, \bibinfo {author}
  {\bibfnamefont {S.}~\bibnamefont {Rajagopalan}}, \bibinfo {author}
  {\bibfnamefont {R.}~\bibnamefont {Stata}}, \bibinfo {author} {\bibfnamefont
  {A.}~\bibnamefont {Tomkins}},\ and\ \bibinfo {author} {\bibfnamefont
  {J.}~\bibnamefont {Wiener}},\ }\bibfield  {title} {\bibinfo {title} {Graph
  structure in the web},\ }\href@noop {} {\bibfield  {journal} {\bibinfo
  {journal} {Computer networks}\ }\textbf {\bibinfo {volume} {33}},\ \bibinfo
  {pages} {309} (\bibinfo {year} {2000})}\BibitemShut {NoStop}%
\bibitem [{\citenamefont {Bonacich}(1972)}]{bonacich1972factoring}%
  \BibitemOpen
  \bibfield  {author} {\bibinfo {author} {\bibfnamefont {P.}~\bibnamefont
  {Bonacich}},\ }\bibfield  {title} {\bibinfo {title} {Factoring and weighting
  approaches to status scores and clique identification},\ }\href@noop {}
  {\bibfield  {journal} {\bibinfo  {journal} {Journal of mathematical
  sociology}\ }\textbf {\bibinfo {volume} {2}},\ \bibinfo {pages} {113}
  (\bibinfo {year} {1972})}\BibitemShut {NoStop}%
\bibitem [{\citenamefont {Restrepo}\ \emph {et~al.}(2006)\citenamefont
  {Restrepo}, \citenamefont {Ott},\ and\ \citenamefont
  {Hunt}}]{restrepo2006characterizing}%
  \BibitemOpen
  \bibfield  {author} {\bibinfo {author} {\bibfnamefont {J.~G.}\ \bibnamefont
  {Restrepo}}, \bibinfo {author} {\bibfnamefont {E.}~\bibnamefont {Ott}},\ and\
  \bibinfo {author} {\bibfnamefont {B.~R.}\ \bibnamefont {Hunt}},\ }\bibfield
  {title} {\bibinfo {title} {Characterizing the dynamical importance of network
  nodes and links},\ }\href@noop {} {\bibfield  {journal} {\bibinfo  {journal}
  {Physical review letters}\ }\textbf {\bibinfo {volume} {97}},\ \bibinfo
  {pages} {094102} (\bibinfo {year} {2006})}\BibitemShut {NoStop}%
\bibitem [{\citenamefont {Martin}\ \emph {et~al.}(2014)\citenamefont {Martin},
  \citenamefont {Zhang},\ and\ \citenamefont
  {Newman}}]{martin2014localization}%
  \BibitemOpen
  \bibfield  {author} {\bibinfo {author} {\bibfnamefont {T.}~\bibnamefont
  {Martin}}, \bibinfo {author} {\bibfnamefont {X.}~\bibnamefont {Zhang}},\ and\
  \bibinfo {author} {\bibfnamefont {M.~E.}\ \bibnamefont {Newman}},\ }\bibfield
   {title} {\bibinfo {title} {Localization and centrality in networks},\
  }\href@noop {} {\bibfield  {journal} {\bibinfo  {journal} {Physical review
  E}\ }\textbf {\bibinfo {volume} {90}},\ \bibinfo {pages} {052808} (\bibinfo
  {year} {2014})}\BibitemShut {NoStop}%
\bibitem [{\citenamefont {Krzakala}\ \emph {et~al.}(2013)\citenamefont
  {Krzakala}, \citenamefont {Moore}, \citenamefont {Mossel}, \citenamefont
  {Neeman}, \citenamefont {Sly}, \citenamefont {Zdeborov{\'a}},\ and\
  \citenamefont {Zhang}}]{krzakala2013spectral}%
  \BibitemOpen
  \bibfield  {author} {\bibinfo {author} {\bibfnamefont {F.}~\bibnamefont
  {Krzakala}}, \bibinfo {author} {\bibfnamefont {C.}~\bibnamefont {Moore}},
  \bibinfo {author} {\bibfnamefont {E.}~\bibnamefont {Mossel}}, \bibinfo
  {author} {\bibfnamefont {J.}~\bibnamefont {Neeman}}, \bibinfo {author}
  {\bibfnamefont {A.}~\bibnamefont {Sly}}, \bibinfo {author} {\bibfnamefont
  {L.}~\bibnamefont {Zdeborov{\'a}}},\ and\ \bibinfo {author} {\bibfnamefont
  {P.}~\bibnamefont {Zhang}},\ }\bibfield  {title} {\bibinfo {title} {Spectral
  redemption in clustering sparse networks},\ }\href@noop {} {\bibfield
  {journal} {\bibinfo  {journal} {Proceedings of the National Academy of
  Sciences}\ }\textbf {\bibinfo {volume} {110}},\ \bibinfo {pages} {20935}
  (\bibinfo {year} {2013})}\BibitemShut {NoStop}%
\bibitem [{\citenamefont {Bordenave}\ \emph {et~al.}(2015)\citenamefont
  {Bordenave}, \citenamefont {Lelarge},\ and\ \citenamefont
  {Massouli{\'e}}}]{bordenave2015non}%
  \BibitemOpen
  \bibfield  {author} {\bibinfo {author} {\bibfnamefont {C.}~\bibnamefont
  {Bordenave}}, \bibinfo {author} {\bibfnamefont {M.}~\bibnamefont {Lelarge}},\
  and\ \bibinfo {author} {\bibfnamefont {L.}~\bibnamefont {Massouli{\'e}}},\
  }\bibfield  {title} {\bibinfo {title} {Non-backtracking spectrum of random
  graphs: community detection and non-regular ramanujan graphs},\ }in\
  \href@noop {} {\emph {\bibinfo {booktitle} {2015 IEEE 56th Annual Symposium
  on Foundations of Computer Science}}}\ (\bibinfo {organization} {IEEE},\
  \bibinfo {year} {2015})\ pp.\ \bibinfo {pages} {1347--1357}\BibitemShut
  {NoStop}%
\bibitem [{\citenamefont {Kawamoto}(2018)}]{kawamoto2018algorithmic}%
  \BibitemOpen
  \bibfield  {author} {\bibinfo {author} {\bibfnamefont {T.}~\bibnamefont
  {Kawamoto}},\ }\bibfield  {title} {\bibinfo {title} {Algorithmic
  detectability threshold of the stochastic block model},\ }\href@noop {}
  {\bibfield  {journal} {\bibinfo  {journal} {Physical Review E}\ }\textbf
  {\bibinfo {volume} {97}},\ \bibinfo {pages} {032301} (\bibinfo {year}
  {2018})}\BibitemShut {NoStop}%
\bibitem [{\citenamefont {Bordenave}\ \emph {et~al.}(2022)\citenamefont
  {Bordenave}, \citenamefont {Coste},\ and\ \citenamefont
  {Nadakuditi}}]{bordenave2022detection}%
  \BibitemOpen
  \bibfield  {author} {\bibinfo {author} {\bibfnamefont {C.}~\bibnamefont
  {Bordenave}}, \bibinfo {author} {\bibfnamefont {S.}~\bibnamefont {Coste}},\
  and\ \bibinfo {author} {\bibfnamefont {R.~R.}\ \bibnamefont {Nadakuditi}},\
  }\bibfield  {title} {\bibinfo {title} {Detection thresholds in very sparse
  matrix completion},\ }\href@noop {} {\bibfield  {journal} {\bibinfo
  {journal} {Foundations of Computational Mathematics}\ ,\ \bibinfo {pages}
  {1}} (\bibinfo {year} {2022})}\BibitemShut {NoStop}%
\bibitem [{\citenamefont {Wasserman}(1980)}]{Wasserman1980AStochastic}%
  \BibitemOpen
  \bibfield  {author} {\bibinfo {author} {\bibfnamefont {S.~S.}\ \bibnamefont
  {Wasserman}},\ }\bibfield  {title} {\bibinfo {title} {A stochastic model for
  directed graphs with transition rates determined by reciprocity},\ }\href
  {http://www.jstor.org/stable/270870} {\bibfield  {journal} {\bibinfo
  {journal} {Sociological Methodology}\ }\textbf {\bibinfo {volume} {11}},\
  \bibinfo {pages} {392} (\bibinfo {year} {1980})}\BibitemShut {NoStop}%
\bibitem [{\citenamefont {Tapias}\ and\ \citenamefont
  {Sollich}(2022)}]{Tapias2022Localization}%
  \BibitemOpen
  \bibfield  {author} {\bibinfo {author} {\bibfnamefont {D.}~\bibnamefont
  {Tapias}}\ and\ \bibinfo {author} {\bibfnamefont {P.}~\bibnamefont
  {Sollich}},\ }\bibfield  {title} {\bibinfo {title} {Localization properties
  of the sparse barrat-m\'ezard trap model},\ }\href
  {https://doi.org/10.1103/PhysRevE.105.054109} {\bibfield  {journal} {\bibinfo
   {journal} {Phys. Rev. E}\ }\textbf {\bibinfo {volume} {105}},\ \bibinfo
  {pages} {054109} (\bibinfo {year} {2022})}\BibitemShut {NoStop}%
\bibitem [{\citenamefont {Hatano}\ and\ \citenamefont
  {Nelson}(1996)}]{HN_PhysRevLett.77.570}%
  \BibitemOpen
  \bibfield  {author} {\bibinfo {author} {\bibfnamefont {N.}~\bibnamefont
  {Hatano}}\ and\ \bibinfo {author} {\bibfnamefont {D.~R.}\ \bibnamefont
  {Nelson}},\ }\bibfield  {title} {\bibinfo {title} {Localization transitions
  in non-hermitian quantum mechanics},\ }\href
  {https://doi.org/10.1103/PhysRevLett.77.570} {\bibfield  {journal} {\bibinfo
  {journal} {Phys. Rev. Lett.}\ }\textbf {\bibinfo {volume} {77}},\ \bibinfo
  {pages} {570} (\bibinfo {year} {1996})}\BibitemShut {NoStop}%
\bibitem [{\citenamefont {Hatano}\ and\ \citenamefont
  {Nelson}(1997)}]{HN_PhysRevB.56.8651}%
  \BibitemOpen
  \bibfield  {author} {\bibinfo {author} {\bibfnamefont {N.}~\bibnamefont
  {Hatano}}\ and\ \bibinfo {author} {\bibfnamefont {D.~R.}\ \bibnamefont
  {Nelson}},\ }\bibfield  {title} {\bibinfo {title} {Vortex pinning and
  non-hermitian quantum mechanics},\ }\href
  {https://doi.org/10.1103/PhysRevB.56.8651} {\bibfield  {journal} {\bibinfo
  {journal} {Phys. Rev. B}\ }\textbf {\bibinfo {volume} {56}},\ \bibinfo
  {pages} {8651} (\bibinfo {year} {1997})}\BibitemShut {NoStop}%
\bibitem [{\citenamefont {Hatano}\ and\ \citenamefont
  {Nelson}(1998)}]{HN_PhysRevB.58.8384}%
  \BibitemOpen
  \bibfield  {author} {\bibinfo {author} {\bibfnamefont {N.}~\bibnamefont
  {Hatano}}\ and\ \bibinfo {author} {\bibfnamefont {D.~R.}\ \bibnamefont
  {Nelson}},\ }\bibfield  {title} {\bibinfo {title} {Non-hermitian
  delocalization and eigenfunctions},\ }\href
  {https://doi.org/10.1103/PhysRevB.58.8384} {\bibfield  {journal} {\bibinfo
  {journal} {Phys. Rev. B}\ }\textbf {\bibinfo {volume} {58}},\ \bibinfo
  {pages} {8384} (\bibinfo {year} {1998})}\BibitemShut {NoStop}%
\bibitem [{\citenamefont {Das~Sarma}\ \emph {et~al.}(1988)\citenamefont
  {Das~Sarma}, \citenamefont {He},\ and\ \citenamefont
  {Xie}}]{DasSarma1988Mobility}%
  \BibitemOpen
  \bibfield  {author} {\bibinfo {author} {\bibfnamefont {S.}~\bibnamefont
  {Das~Sarma}}, \bibinfo {author} {\bibfnamefont {S.}~\bibnamefont {He}},\ and\
  \bibinfo {author} {\bibfnamefont {X.~C.}\ \bibnamefont {Xie}},\ }\bibfield
  {title} {\bibinfo {title} {Mobility edge in a model one-dimensional
  potential},\ }\href {https://doi.org/10.1103/PhysRevLett.61.2144} {\bibfield
  {journal} {\bibinfo  {journal} {Phys. Rev. Lett.}\ }\textbf {\bibinfo
  {volume} {61}},\ \bibinfo {pages} {2144} (\bibinfo {year}
  {1988})}\BibitemShut {NoStop}%
\bibitem [{\citenamefont {Das~Sarma}\ \emph {et~al.}(1990)\citenamefont
  {Das~Sarma}, \citenamefont {He},\ and\ \citenamefont
  {Xie}}]{DasSarma1990Localization}%
  \BibitemOpen
  \bibfield  {author} {\bibinfo {author} {\bibfnamefont {S.}~\bibnamefont
  {Das~Sarma}}, \bibinfo {author} {\bibfnamefont {S.}~\bibnamefont {He}},\ and\
  \bibinfo {author} {\bibfnamefont {X.~C.}\ \bibnamefont {Xie}},\ }\bibfield
  {title} {\bibinfo {title} {Localization, mobility edges, and metal-insulator
  transition in a class of one-dimensional slowly varying deterministic
  potentials},\ }\href {https://doi.org/10.1103/PhysRevB.41.5544} {\bibfield
  {journal} {\bibinfo  {journal} {Phys. Rev. B}\ }\textbf {\bibinfo {volume}
  {41}},\ \bibinfo {pages} {5544} (\bibinfo {year} {1990})}\BibitemShut
  {NoStop}%
\bibitem [{\citenamefont {Liu}\ \emph {et~al.}(2020)\citenamefont {Liu},
  \citenamefont {Guo}, \citenamefont {Pu},\ and\ \citenamefont
  {Longhi}}]{Liu2020Generalized}%
  \BibitemOpen
  \bibfield  {author} {\bibinfo {author} {\bibfnamefont {T.}~\bibnamefont
  {Liu}}, \bibinfo {author} {\bibfnamefont {H.}~\bibnamefont {Guo}}, \bibinfo
  {author} {\bibfnamefont {Y.}~\bibnamefont {Pu}},\ and\ \bibinfo {author}
  {\bibfnamefont {S.}~\bibnamefont {Longhi}},\ }\bibfield  {title} {\bibinfo
  {title} {Generalized aubry-andr\'e self-duality and mobility edges in
  non-hermitian quasiperiodic lattices},\ }\href
  {https://doi.org/10.1103/PhysRevB.102.024205} {\bibfield  {journal} {\bibinfo
   {journal} {Phys. Rev. B}\ }\textbf {\bibinfo {volume} {102}},\ \bibinfo
  {pages} {024205} (\bibinfo {year} {2020})}\BibitemShut {NoStop}%
\bibitem [{\citenamefont {Peron}\ \emph {et~al.}(2020)\citenamefont {Peron},
  \citenamefont {de~Resende}, \citenamefont {Rodrigues}, \citenamefont
  {Costa},\ and\ \citenamefont {M\'endez-Berm\'udez}}]{peron2020spacing}%
  \BibitemOpen
  \bibfield  {author} {\bibinfo {author} {\bibfnamefont {T.}~\bibnamefont
  {Peron}}, \bibinfo {author} {\bibfnamefont {B.~M.~F.}\ \bibnamefont
  {de~Resende}}, \bibinfo {author} {\bibfnamefont {F.~A.}\ \bibnamefont
  {Rodrigues}}, \bibinfo {author} {\bibfnamefont {L.~d.~F.}\ \bibnamefont
  {Costa}},\ and\ \bibinfo {author} {\bibfnamefont {J.~A.}\ \bibnamefont
  {M\'endez-Berm\'udez}},\ }\bibfield  {title} {\bibinfo {title} {Spacing ratio
  characterization of the spectra of directed random networks},\ }\href
  {https://doi.org/10.1103/PhysRevE.102.062305} {\bibfield  {journal} {\bibinfo
   {journal} {Phys. Rev. E}\ }\textbf {\bibinfo {volume} {102}},\ \bibinfo
  {pages} {062305} (\bibinfo {year} {2020})}\BibitemShut {NoStop}%
\bibitem [{\citenamefont {S\'a}\ \emph {et~al.}(2020)\citenamefont {S\'a},
  \citenamefont {Ribeiro},\ and\ \citenamefont {Prosen}}]{sa2020complex}%
  \BibitemOpen
  \bibfield  {author} {\bibinfo {author} {\bibfnamefont {L.}~\bibnamefont
  {S\'a}}, \bibinfo {author} {\bibfnamefont {P.}~\bibnamefont {Ribeiro}},\ and\
  \bibinfo {author} {\bibfnamefont {T.~c.~v.}\ \bibnamefont {Prosen}},\
  }\bibfield  {title} {\bibinfo {title} {Complex spacing ratios: A signature of
  dissipative quantum chaos},\ }\href
  {https://doi.org/10.1103/PhysRevX.10.021019} {\bibfield  {journal} {\bibinfo
  {journal} {Phys. Rev. X}\ }\textbf {\bibinfo {volume} {10}},\ \bibinfo
  {pages} {021019} (\bibinfo {year} {2020})}\BibitemShut {NoStop}%
\bibitem [{\citenamefont {Hamazaki}\ \emph {et~al.}(2019)\citenamefont
  {Hamazaki}, \citenamefont {Kawabata},\ and\ \citenamefont
  {Ueda}}]{Hamazaki2019NonHermitian}%
  \BibitemOpen
  \bibfield  {author} {\bibinfo {author} {\bibfnamefont {R.}~\bibnamefont
  {Hamazaki}}, \bibinfo {author} {\bibfnamefont {K.}~\bibnamefont {Kawabata}},\
  and\ \bibinfo {author} {\bibfnamefont {M.}~\bibnamefont {Ueda}},\ }\bibfield
  {title} {\bibinfo {title} {Non-hermitian many-body localization},\ }\href
  {https://doi.org/10.1103/PhysRevLett.123.090603} {\bibfield  {journal}
  {\bibinfo  {journal} {Phys. Rev. Lett.}\ }\textbf {\bibinfo {volume} {123}},\
  \bibinfo {pages} {090603} (\bibinfo {year} {2019})}\BibitemShut {NoStop}%
\bibitem [{\citenamefont {Huang}\ and\ \citenamefont
  {Shklovskii}(2020)}]{Huang2020Anderson}%
  \BibitemOpen
  \bibfield  {author} {\bibinfo {author} {\bibfnamefont {Y.}~\bibnamefont
  {Huang}}\ and\ \bibinfo {author} {\bibfnamefont {B.~I.}\ \bibnamefont
  {Shklovskii}},\ }\bibfield  {title} {\bibinfo {title} {Anderson transition in
  three-dimensional systems with non-hermitian disorder},\ }\href
  {https://doi.org/10.1103/PhysRevB.101.014204} {\bibfield  {journal} {\bibinfo
   {journal} {Phys. Rev. B}\ }\textbf {\bibinfo {volume} {101}},\ \bibinfo
  {pages} {014204} (\bibinfo {year} {2020})}\BibitemShut {NoStop}%
\bibitem [{\citenamefont {Tzortzakakis}\ \emph {et~al.}(2020)\citenamefont
  {Tzortzakakis}, \citenamefont {Makris},\ and\ \citenamefont
  {Economou}}]{Tzortzakakis2020NonHermitian}%
  \BibitemOpen
  \bibfield  {author} {\bibinfo {author} {\bibfnamefont {A.~F.}\ \bibnamefont
  {Tzortzakakis}}, \bibinfo {author} {\bibfnamefont {K.~G.}\ \bibnamefont
  {Makris}},\ and\ \bibinfo {author} {\bibfnamefont {E.~N.}\ \bibnamefont
  {Economou}},\ }\bibfield  {title} {\bibinfo {title} {Non-hermitian disorder
  in two-dimensional optical lattices},\ }\href
  {https://doi.org/10.1103/PhysRevB.101.014202} {\bibfield  {journal} {\bibinfo
   {journal} {Phys. Rev. B}\ }\textbf {\bibinfo {volume} {101}},\ \bibinfo
  {pages} {014202} (\bibinfo {year} {2020})}\BibitemShut {NoStop}%
\bibitem [{\citenamefont {Luo}\ \emph {et~al.}(2021)\citenamefont {Luo},
  \citenamefont {Ohtsuki},\ and\ \citenamefont
  {Shindou}}]{Luo2021Universality}%
  \BibitemOpen
  \bibfield  {author} {\bibinfo {author} {\bibfnamefont {X.}~\bibnamefont
  {Luo}}, \bibinfo {author} {\bibfnamefont {T.}~\bibnamefont {Ohtsuki}},\ and\
  \bibinfo {author} {\bibfnamefont {R.}~\bibnamefont {Shindou}},\ }\bibfield
  {title} {\bibinfo {title} {Universality classes of the anderson transitions
  driven by non-hermitian disorder},\ }\href
  {https://doi.org/10.1103/PhysRevLett.126.090402} {\bibfield  {journal}
  {\bibinfo  {journal} {Phys. Rev. Lett.}\ }\textbf {\bibinfo {volume} {126}},\
  \bibinfo {pages} {090402} (\bibinfo {year} {2021})}\BibitemShut {NoStop}%
\bibitem [{\citenamefont {De~Tomasi}\ and\ \citenamefont
  {Khaymovich}(2022)}]{DeTomasi2022NonHermitian}%
  \BibitemOpen
  \bibfield  {author} {\bibinfo {author} {\bibfnamefont {G.}~\bibnamefont
  {De~Tomasi}}\ and\ \bibinfo {author} {\bibfnamefont {I.~M.}\ \bibnamefont
  {Khaymovich}},\ }\bibfield  {title} {\bibinfo {title} {Non-hermitian
  rosenzweig-porter random-matrix ensemble: Obstruction to the fractal phase},\
  }\href {https://doi.org/10.1103/PhysRevB.106.094204} {\bibfield  {journal}
  {\bibinfo  {journal} {Phys. Rev. B}\ }\textbf {\bibinfo {volume} {106}},\
  \bibinfo {pages} {094204} (\bibinfo {year} {2022})}\BibitemShut {NoStop}%
\bibitem [{\citenamefont {De~Tomasi}\ and\ \citenamefont
  {Khaymovich}(2023)}]{detomasi2023nonhermiticity}%
  \BibitemOpen
  \bibfield  {author} {\bibinfo {author} {\bibfnamefont {G.}~\bibnamefont
  {De~Tomasi}}\ and\ \bibinfo {author} {\bibfnamefont {I.~M.}\ \bibnamefont
  {Khaymovich}},\ }\bibfield  {title} {\bibinfo {title} {Non-hermiticity
  induces localization: Good and bad resonances in power-law random banded
  matrices},\ }\href {https://doi.org/10.1103/PhysRevB.108.L180202} {\bibfield
  {journal} {\bibinfo  {journal} {Phys. Rev. B}\ }\textbf {\bibinfo {volume}
  {108}},\ \bibinfo {pages} {L180202} (\bibinfo {year} {2023})}\BibitemShut
  {NoStop}%
\bibitem [{\citenamefont {Neri}\ and\ \citenamefont
  {Metz}(2012)}]{Neri2012Spectra}%
  \BibitemOpen
  \bibfield  {author} {\bibinfo {author} {\bibfnamefont {I.}~\bibnamefont
  {Neri}}\ and\ \bibinfo {author} {\bibfnamefont {F.~L.}\ \bibnamefont
  {Metz}},\ }\bibfield  {title} {\bibinfo {title} {Spectra of sparse
  non-hermitian random matrices: An analytical solution},\ }\href
  {https://doi.org/10.1103/PhysRevLett.109.030602} {\bibfield  {journal}
  {\bibinfo  {journal} {Phys. Rev. Lett.}\ }\textbf {\bibinfo {volume} {109}},\
  \bibinfo {pages} {030602} (\bibinfo {year} {2012})}\BibitemShut {NoStop}%
\bibitem [{\citenamefont {Neri}\ and\ \citenamefont
  {Metz}(2016)}]{Neri2016Eigenvalue}%
  \BibitemOpen
  \bibfield  {author} {\bibinfo {author} {\bibfnamefont {I.}~\bibnamefont
  {Neri}}\ and\ \bibinfo {author} {\bibfnamefont {F.~L.}\ \bibnamefont
  {Metz}},\ }\bibfield  {title} {\bibinfo {title} {Eigenvalue outliers of
  non-hermitian random matrices with a local tree structure},\ }\href
  {https://doi.org/10.1103/PhysRevLett.117.224101} {\bibfield  {journal}
  {\bibinfo  {journal} {Phys. Rev. Lett.}\ }\textbf {\bibinfo {volume} {117}},\
  \bibinfo {pages} {224101} (\bibinfo {year} {2016})}\BibitemShut {NoStop}%
\bibitem [{\citenamefont {Metz}\ \emph {et~al.}(2019)\citenamefont {Metz},
  \citenamefont {Neri},\ and\ \citenamefont {Rogers}}]{Metz2019Spectral}%
  \BibitemOpen
  \bibfield  {author} {\bibinfo {author} {\bibfnamefont {F.~L.}\ \bibnamefont
  {Metz}}, \bibinfo {author} {\bibfnamefont {I.}~\bibnamefont {Neri}},\ and\
  \bibinfo {author} {\bibfnamefont {T.}~\bibnamefont {Rogers}},\ }\bibfield
  {title} {\bibinfo {title} {Spectral theory of sparse non-hermitian random
  matrices},\ }\href {https://doi.org/10.1088/1751-8121/ab1ce0} {\bibfield
  {journal} {\bibinfo  {journal} {Journal of Physics A: Mathematical and
  Theoretical}\ }\textbf {\bibinfo {volume} {52}},\ \bibinfo {pages} {434003}
  (\bibinfo {year} {2019})}\BibitemShut {NoStop}%
\bibitem [{\citenamefont {Metz}\ and\ \citenamefont
  {Neri}(2021)}]{Metz2021Localization}%
  \BibitemOpen
  \bibfield  {author} {\bibinfo {author} {\bibfnamefont {F.~L.}\ \bibnamefont
  {Metz}}\ and\ \bibinfo {author} {\bibfnamefont {I.}~\bibnamefont {Neri}},\
  }\bibfield  {title} {\bibinfo {title} {Localization and universality of
  eigenvectors in directed random graphs},\ }\href
  {https://doi.org/10.1103/PhysRevLett.126.040604} {\bibfield  {journal}
  {\bibinfo  {journal} {Phys. Rev. Lett.}\ }\textbf {\bibinfo {volume} {126}},\
  \bibinfo {pages} {040604} (\bibinfo {year} {2021})}\BibitemShut {NoStop}%
\bibitem [{\citenamefont {Kunst}\ \emph {et~al.}(2018)\citenamefont {Kunst},
  \citenamefont {Edvardsson}, \citenamefont {Budich},\ and\ \citenamefont
  {Bergholtz}}]{Kunst2018Biorthogonal}%
  \BibitemOpen
  \bibfield  {author} {\bibinfo {author} {\bibfnamefont {F.~K.}\ \bibnamefont
  {Kunst}}, \bibinfo {author} {\bibfnamefont {E.}~\bibnamefont {Edvardsson}},
  \bibinfo {author} {\bibfnamefont {J.~C.}\ \bibnamefont {Budich}},\ and\
  \bibinfo {author} {\bibfnamefont {E.~J.}\ \bibnamefont {Bergholtz}},\
  }\bibfield  {title} {\bibinfo {title} {Biorthogonal bulk-boundary
  correspondence in non-hermitian systems},\ }\href
  {https://doi.org/10.1103/PhysRevLett.121.026808} {\bibfield  {journal}
  {\bibinfo  {journal} {Phys. Rev. Lett.}\ }\textbf {\bibinfo {volume} {121}},\
  \bibinfo {pages} {026808} (\bibinfo {year} {2018})}\BibitemShut {NoStop}%
\bibitem [{\citenamefont {Yao}\ and\ \citenamefont {Wang}(2018)}]{Yao2018Edge}%
  \BibitemOpen
  \bibfield  {author} {\bibinfo {author} {\bibfnamefont {S.}~\bibnamefont
  {Yao}}\ and\ \bibinfo {author} {\bibfnamefont {Z.}~\bibnamefont {Wang}},\
  }\bibfield  {title} {\bibinfo {title} {Edge states and topological invariants
  of non-hermitian systems},\ }\href
  {https://doi.org/10.1103/PhysRevLett.121.086803} {\bibfield  {journal}
  {\bibinfo  {journal} {Phys. Rev. Lett.}\ }\textbf {\bibinfo {volume} {121}},\
  \bibinfo {pages} {086803} (\bibinfo {year} {2018})}\BibitemShut {NoStop}%
\bibitem [{\citenamefont {Xiong}(2018)}]{Xiong2018Why}%
  \BibitemOpen
  \bibfield  {author} {\bibinfo {author} {\bibfnamefont {Y.}~\bibnamefont
  {Xiong}},\ }\bibfield  {title} {\bibinfo {title} {Why does bulk boundary
  correspondence fail in some non-hermitian topological models},\ }\href
  {https://doi.org/10.1088/2399-6528/aab64a} {\bibfield  {journal} {\bibinfo
  {journal} {Journal of Physics Communications}\ }\textbf {\bibinfo {volume}
  {2}},\ \bibinfo {pages} {035043} (\bibinfo {year} {2018})}\BibitemShut
  {NoStop}%
\bibitem [{\citenamefont {Kawabata}\ \emph {et~al.}(2023)\citenamefont
  {Kawabata}, \citenamefont {Numasawa},\ and\ \citenamefont
  {Ryu}}]{Kawabata2023Entanglement}%
  \BibitemOpen
  \bibfield  {author} {\bibinfo {author} {\bibfnamefont {K.}~\bibnamefont
  {Kawabata}}, \bibinfo {author} {\bibfnamefont {T.}~\bibnamefont {Numasawa}},\
  and\ \bibinfo {author} {\bibfnamefont {S.}~\bibnamefont {Ryu}},\ }\bibfield
  {title} {\bibinfo {title} {Entanglement phase transition induced by the
  non-hermitian skin effect},\ }\href
  {https://doi.org/10.1103/PhysRevX.13.021007} {\bibfield  {journal} {\bibinfo
  {journal} {Phys. Rev. X}\ }\textbf {\bibinfo {volume} {13}},\ \bibinfo
  {pages} {021007} (\bibinfo {year} {2023})}\BibitemShut {NoStop}%
\bibitem [{\citenamefont {Bergholtz}\ \emph {et~al.}(2021)\citenamefont
  {Bergholtz}, \citenamefont {Budich},\ and\ \citenamefont
  {Kunst}}]{Bergholtz2021Exceptional}%
  \BibitemOpen
  \bibfield  {author} {\bibinfo {author} {\bibfnamefont {E.~J.}\ \bibnamefont
  {Bergholtz}}, \bibinfo {author} {\bibfnamefont {J.~C.}\ \bibnamefont
  {Budich}},\ and\ \bibinfo {author} {\bibfnamefont {F.~K.}\ \bibnamefont
  {Kunst}},\ }\bibfield  {title} {\bibinfo {title} {Exceptional topology of
  non-hermitian systems},\ }\href
  {https://doi.org/10.1103/RevModPhys.93.015005} {\bibfield  {journal}
  {\bibinfo  {journal} {Rev. Mod. Phys.}\ }\textbf {\bibinfo {volume} {93}},\
  \bibinfo {pages} {015005} (\bibinfo {year} {2021})}\BibitemShut {NoStop}%
\bibitem [{\citenamefont {Okuma}\ and\ \citenamefont
  {Sato}(2023)}]{Okuma2023NonHermitian}%
  \BibitemOpen
  \bibfield  {author} {\bibinfo {author} {\bibfnamefont {N.}~\bibnamefont
  {Okuma}}\ and\ \bibinfo {author} {\bibfnamefont {M.}~\bibnamefont {Sato}},\
  }\bibfield  {title} {\bibinfo {title} {Non-hermitian topological phenomena: A
  review},\ }\href {https://doi.org/10.1146/annurev-conmatphys-040521-033133}
  {\bibfield  {journal} {\bibinfo  {journal} {Annual Review of Condensed Matter
  Physics}\ }\textbf {\bibinfo {volume} {14}},\ \bibinfo {pages} {83} (\bibinfo
  {year} {2023})},\ \Eprint
  {https://arxiv.org/abs/https://doi.org/10.1146/annurev-conmatphys-040521-033133}
  {https://doi.org/10.1146/annurev-conmatphys-040521-033133} \BibitemShut
  {NoStop}%
\bibitem [{\citenamefont {Brody}(2013)}]{Brody2014Biorthogonal}%
  \BibitemOpen
  \bibfield  {author} {\bibinfo {author} {\bibfnamefont {D.~C.}\ \bibnamefont
  {Brody}},\ }\bibfield  {title} {\bibinfo {title} {Biorthogonal quantum
  mechanics},\ }\href {https://doi.org/10.1088/1751-8113/47/3/035305}
  {\bibfield  {journal} {\bibinfo  {journal} {Journal of Physics A:
  Mathematical and Theoretical}\ }\textbf {\bibinfo {volume} {47}},\ \bibinfo
  {pages} {035305} (\bibinfo {year} {2013})}\BibitemShut {NoStop}%
\bibitem [{\citenamefont {Ashida}\ \emph {et~al.}(2020)\citenamefont {Ashida},
  \citenamefont {Gong},\ and\ \citenamefont {Ueda}}]{Ashida2020NonHermitian}%
  \BibitemOpen
  \bibfield  {author} {\bibinfo {author} {\bibfnamefont {Y.}~\bibnamefont
  {Ashida}}, \bibinfo {author} {\bibfnamefont {Z.}~\bibnamefont {Gong}},\ and\
  \bibinfo {author} {\bibfnamefont {M.}~\bibnamefont {Ueda}},\ }\bibfield
  {title} {\bibinfo {title} {Non-hermitian physics},\ }\href
  {https://doi.org/10.1080/00018732.2021.1876991} {\bibfield  {journal}
  {\bibinfo  {journal} {Advances in Physics}\ }\textbf {\bibinfo {volume}
  {69}},\ \bibinfo {pages} {249} (\bibinfo {year} {2020})},\ \Eprint
  {https://arxiv.org/abs/https://doi.org/10.1080/00018732.2021.1876991}
  {https://doi.org/10.1080/00018732.2021.1876991} \BibitemShut {NoStop}%
\bibitem [{\citenamefont {Gong}\ \emph {et~al.}(2018)\citenamefont {Gong},
  \citenamefont {Ashida}, \citenamefont {Kawabata}, \citenamefont {Takasan},
  \citenamefont {Higashikawa},\ and\ \citenamefont
  {Ueda}}]{Gong2018Topological}%
  \BibitemOpen
  \bibfield  {author} {\bibinfo {author} {\bibfnamefont {Z.}~\bibnamefont
  {Gong}}, \bibinfo {author} {\bibfnamefont {Y.}~\bibnamefont {Ashida}},
  \bibinfo {author} {\bibfnamefont {K.}~\bibnamefont {Kawabata}}, \bibinfo
  {author} {\bibfnamefont {K.}~\bibnamefont {Takasan}}, \bibinfo {author}
  {\bibfnamefont {S.}~\bibnamefont {Higashikawa}},\ and\ \bibinfo {author}
  {\bibfnamefont {M.}~\bibnamefont {Ueda}},\ }\bibfield  {title} {\bibinfo
  {title} {Topological phases of non-hermitian systems},\ }\href
  {https://doi.org/10.1103/PhysRevX.8.031079} {\bibfield  {journal} {\bibinfo
  {journal} {Phys. Rev. X}\ }\textbf {\bibinfo {volume} {8}},\ \bibinfo {pages}
  {031079} (\bibinfo {year} {2018})}\BibitemShut {NoStop}%
\bibitem [{\citenamefont {Xiao}\ and\ \citenamefont
  {Chan}(2022)}]{Xiao2022Topology}%
  \BibitemOpen
  \bibfield  {author} {\bibinfo {author} {\bibfnamefont {Y.-X.}\ \bibnamefont
  {Xiao}}\ and\ \bibinfo {author} {\bibfnamefont {C.~T.}\ \bibnamefont
  {Chan}},\ }\bibfield  {title} {\bibinfo {title} {Topology in non-hermitian
  chern insulators with skin effect},\ }\href
  {https://doi.org/10.1103/PhysRevB.105.075128} {\bibfield  {journal} {\bibinfo
   {journal} {Phys. Rev. B}\ }\textbf {\bibinfo {volume} {105}},\ \bibinfo
  {pages} {075128} (\bibinfo {year} {2022})}\BibitemShut {NoStop}%
\bibitem [{\citenamefont {Suthar}\ \emph {et~al.}(2022)\citenamefont {Suthar},
  \citenamefont {Wang}, \citenamefont {Huang}, \citenamefont {Jen},\ and\
  \citenamefont {You}}]{Suthar2022NonHermitian}%
  \BibitemOpen
  \bibfield  {author} {\bibinfo {author} {\bibfnamefont {K.}~\bibnamefont
  {Suthar}}, \bibinfo {author} {\bibfnamefont {Y.-C.}\ \bibnamefont {Wang}},
  \bibinfo {author} {\bibfnamefont {Y.-P.}\ \bibnamefont {Huang}}, \bibinfo
  {author} {\bibfnamefont {H.~H.}\ \bibnamefont {Jen}},\ and\ \bibinfo {author}
  {\bibfnamefont {J.-S.}\ \bibnamefont {You}},\ }\bibfield  {title} {\bibinfo
  {title} {Non-hermitian many-body localization with open boundaries},\ }\href
  {https://doi.org/10.1103/PhysRevB.106.064208} {\bibfield  {journal} {\bibinfo
   {journal} {Phys. Rev. B}\ }\textbf {\bibinfo {volume} {106}},\ \bibinfo
  {pages} {064208} (\bibinfo {year} {2022})}\BibitemShut {NoStop}%
\bibitem [{\citenamefont {Moudgalya}\ \emph {et~al.}(2022)\citenamefont
  {Moudgalya}, \citenamefont {Bernevig},\ and\ \citenamefont
  {Regnault}}]{Moudgalya2022Quantum}%
  \BibitemOpen
  \bibfield  {author} {\bibinfo {author} {\bibfnamefont {S.}~\bibnamefont
  {Moudgalya}}, \bibinfo {author} {\bibfnamefont {B.~A.}\ \bibnamefont
  {Bernevig}},\ and\ \bibinfo {author} {\bibfnamefont {N.}~\bibnamefont
  {Regnault}},\ }\bibfield  {title} {\bibinfo {title} {Quantum many-body scars
  and hilbert space fragmentation: a review of exact results},\ }\href
  {https://doi.org/10.1088/1361-6633/ac73a0} {\bibfield  {journal} {\bibinfo
  {journal} {Reports on Progress in Physics}\ }\textbf {\bibinfo {volume}
  {85}},\ \bibinfo {pages} {086501} (\bibinfo {year} {2022})}\BibitemShut
  {NoStop}%
\bibitem [{\citenamefont {Tikhonov}\ \emph {et~al.}(2016)\citenamefont
  {Tikhonov}, \citenamefont {Mirlin},\ and\ \citenamefont
  {Skvortsov}}]{Tikhonov2016Anderson}%
  \BibitemOpen
  \bibfield  {author} {\bibinfo {author} {\bibfnamefont {K.~S.}\ \bibnamefont
  {Tikhonov}}, \bibinfo {author} {\bibfnamefont {A.~D.}\ \bibnamefont
  {Mirlin}},\ and\ \bibinfo {author} {\bibfnamefont {M.~A.}\ \bibnamefont
  {Skvortsov}},\ }\bibfield  {title} {\bibinfo {title} {Anderson localization
  and ergodicity on random regular graphs},\ }\href
  {https://doi.org/10.1103/PhysRevB.94.220203} {\bibfield  {journal} {\bibinfo
  {journal} {Phys. Rev. B}\ }\textbf {\bibinfo {volume} {94}},\ \bibinfo
  {pages} {220203} (\bibinfo {year} {2016})}\BibitemShut {NoStop}%
\bibitem [{\citenamefont {De~Luca}\ \emph {et~al.}(2014)\citenamefont
  {De~Luca}, \citenamefont {Altshuler}, \citenamefont {Kravtsov},\ and\
  \citenamefont {Scardicchio}}]{Luca2014}%
  \BibitemOpen
  \bibfield  {author} {\bibinfo {author} {\bibfnamefont {A.}~\bibnamefont
  {De~Luca}}, \bibinfo {author} {\bibfnamefont {B.}~\bibnamefont {Altshuler}},
  \bibinfo {author} {\bibfnamefont {V.}~\bibnamefont {Kravtsov}},\ and\
  \bibinfo {author} {\bibfnamefont {A.}~\bibnamefont {Scardicchio}},\
  }\bibfield  {title} {\bibinfo {title} {{{Anderson}} localization on the
  {Bethe} lattice: Nonergodicity of extended states},\ }\href
  {https://doi.org/10.1103/PhysRevLett.113.046806} {\bibfield  {journal}
  {\bibinfo  {journal} {Phys. Rev. Lett.}\ }\textbf {\bibinfo {volume} {113}},\
  \bibinfo {pages} {046806} (\bibinfo {year} {2014})}\BibitemShut {NoStop}%
\bibitem [{\citenamefont {Erd\H{o}s}\ and\ \citenamefont {R\'enyi}(1959)}]{ER}%
  \BibitemOpen
  \bibfield  {author} {\bibinfo {author} {\bibfnamefont {P.}~\bibnamefont
  {Erd\H{o}s}}\ and\ \bibinfo {author} {\bibfnamefont {A.}~\bibnamefont
  {R\'enyi}},\ }\bibfield  {title} {\bibinfo {title} {On random graphs},\
  }\href@noop {} {\bibfield  {journal} {\bibinfo  {journal} {Publicationes
  Mathematicae (Debrecen)}\ }\textbf {\bibinfo {volume} {6}},\ \bibinfo {pages}
  {290} (\bibinfo {year} {1959})}\BibitemShut {NoStop}%
\bibitem [{\citenamefont {Barab\'asi}\ and\ \citenamefont {Albert}(1999)}]{BA}%
  \BibitemOpen
  \bibfield  {author} {\bibinfo {author} {\bibfnamefont {A.-L.}\ \bibnamefont
  {Barab\'asi}}\ and\ \bibinfo {author} {\bibfnamefont {R.}~\bibnamefont
  {Albert}},\ }\bibfield  {title} {\bibinfo {title} {Emergence of scaling in
  random networks},\ }\href {https://doi.org/10.1126/science.286.5439.509}
  {\bibfield  {journal} {\bibinfo  {journal} {Science}\ }\textbf {\bibinfo
  {volume} {286}},\ \bibinfo {pages} {509} (\bibinfo {year} {1999})},\ \Eprint
  {https://arxiv.org/abs/https://www.science.org/doi/pdf/10.1126/science.286.5439.509}
  {https://www.science.org/doi/pdf/10.1126/science.286.5439.509} \BibitemShut
  {NoStop}%
\bibitem [{\citenamefont {Watts}\ and\ \citenamefont {Strogatz}(1998)}]{WS}%
  \BibitemOpen
  \bibfield  {author} {\bibinfo {author} {\bibfnamefont {D.~J.}\ \bibnamefont
  {Watts}}\ and\ \bibinfo {author} {\bibfnamefont {S.~H.}\ \bibnamefont
  {Strogatz}},\ }\bibfield  {title} {\bibinfo {title} {Collective dynamics of
  ‘small-world’ networks},\ }\href {https://doi.org/10.1038/30918}
  {\bibfield  {journal} {\bibinfo  {journal} {Nature}\ }\textbf {\bibinfo
  {volume} {393}},\ \bibinfo {pages} {440} (\bibinfo {year}
  {1998})}\BibitemShut {NoStop}%
\bibitem [{\citenamefont {Palla}\ \emph {et~al.}(2007)\citenamefont {Palla},
  \citenamefont {Derényi},\ and\ \citenamefont {Vicsek}}]{Palla2007}%
  \BibitemOpen
  \bibfield  {author} {\bibinfo {author} {\bibfnamefont {G.}~\bibnamefont
  {Palla}}, \bibinfo {author} {\bibfnamefont {I.}~\bibnamefont {Derényi}},\
  and\ \bibinfo {author} {\bibfnamefont {T.}~\bibnamefont {Vicsek}},\
  }\bibfield  {title} {\bibinfo {title} {The critical point of k -clique
  percolation in the erdős–rényi graph},\ }\href
  {https://doi.org/10.1007/s10955-006-9184-x} {\bibfield  {journal} {\bibinfo
  {journal} {Journal of Statistical Physics}\ }\textbf {\bibinfo {volume}
  {128}},\ \bibinfo {pages} {219} (\bibinfo {year} {2007})}\BibitemShut
  {NoStop}%
\bibitem [{\citenamefont {Anderson}(1958)}]{anderson1958}%
  \BibitemOpen
  \bibfield  {author} {\bibinfo {author} {\bibfnamefont {P.~W.}\ \bibnamefont
  {Anderson}},\ }\bibfield  {title} {\bibinfo {title} {Absence of diffusion in
  certain random lattices},\ }\href@noop {} {\bibfield  {journal} {\bibinfo
  {journal} {Physical review}\ }\textbf {\bibinfo {volume} {109}},\ \bibinfo
  {pages} {1492} (\bibinfo {year} {1958})}\BibitemShut {NoStop}%
\bibitem [{\citenamefont {Luo}\ \emph {et~al.}(2022)\citenamefont {Luo},
  \citenamefont {Xiao}, \citenamefont {Kawabata}, \citenamefont {Ohtsuki},\
  and\ \citenamefont {Shindou}}]{Luo2022Unifying}%
  \BibitemOpen
  \bibfield  {author} {\bibinfo {author} {\bibfnamefont {X.}~\bibnamefont
  {Luo}}, \bibinfo {author} {\bibfnamefont {Z.}~\bibnamefont {Xiao}}, \bibinfo
  {author} {\bibfnamefont {K.}~\bibnamefont {Kawabata}}, \bibinfo {author}
  {\bibfnamefont {T.}~\bibnamefont {Ohtsuki}},\ and\ \bibinfo {author}
  {\bibfnamefont {R.}~\bibnamefont {Shindou}},\ }\bibfield  {title} {\bibinfo
  {title} {Unifying the anderson transitions in hermitian and non-hermitian
  systems},\ }\href {https://doi.org/10.1103/PhysRevResearch.4.L022035}
  {\bibfield  {journal} {\bibinfo  {journal} {Phys. Rev. Res.}\ }\textbf
  {\bibinfo {volume} {4}},\ \bibinfo {pages} {L022035} (\bibinfo {year}
  {2022})}\BibitemShut {NoStop}%
\bibitem [{\citenamefont {Kravtsov}\ \emph {et~al.}(2018)\citenamefont
  {Kravtsov}, \citenamefont {Altshuler},\ and\ \citenamefont
  {Ioffe}}]{Kravtsov2018nonergodic}%
  \BibitemOpen
  \bibfield  {author} {\bibinfo {author} {\bibfnamefont {V.~E.}\ \bibnamefont
  {Kravtsov}}, \bibinfo {author} {\bibfnamefont {B.~L.}\ \bibnamefont
  {Altshuler}},\ and\ \bibinfo {author} {\bibfnamefont {L.~B.}\ \bibnamefont
  {Ioffe}},\ }\bibfield  {title} {\bibinfo {title} {Non-ergodic delocalized
  phase in {A}nderson model on {Bethe} lattice and regular graph},\ }\href
  {https://doi.org/https://doi.org/10.1016/j.aop.2017.12.009} {\bibfield
  {journal} {\bibinfo  {journal} {Annals of Physics}\ }\textbf {\bibinfo
  {volume} {389}},\ \bibinfo {pages} {148 } (\bibinfo {year}
  {2018})}\BibitemShut {NoStop}%
\bibitem [{\citenamefont {Parisi}\ \emph {et~al.}(2019)\citenamefont {Parisi},
  \citenamefont {Pascazio}, \citenamefont {Pietracaprina}, \citenamefont
  {Ros},\ and\ \citenamefont {Scardicchio}}]{Parisi2019anderson}%
  \BibitemOpen
  \bibfield  {author} {\bibinfo {author} {\bibfnamefont {G.}~\bibnamefont
  {Parisi}}, \bibinfo {author} {\bibfnamefont {S.}~\bibnamefont {Pascazio}},
  \bibinfo {author} {\bibfnamefont {F.}~\bibnamefont {Pietracaprina}}, \bibinfo
  {author} {\bibfnamefont {V.}~\bibnamefont {Ros}},\ and\ \bibinfo {author}
  {\bibfnamefont {A.}~\bibnamefont {Scardicchio}},\ }\bibfield  {title}
  {\bibinfo {title} {{Anderson} transition on the {Bethe} lattice: an approach
  with real energies},\ }\href {https://doi.org/10.1088/1751-8121/ab56e8}
  {\bibfield  {journal} {\bibinfo  {journal} {Journal of Physics A:
  Mathematical and Theoretical}\ }\textbf {\bibinfo {volume} {53}},\ \bibinfo
  {pages} {014003} (\bibinfo {year} {2019})}\BibitemShut {NoStop}%
\bibitem [{\citenamefont {Scardicchio}\ and\ \citenamefont
  {Thiery}(2017)}]{scardicchio2017perturbation}%
  \BibitemOpen
  \bibfield  {author} {\bibinfo {author} {\bibfnamefont {A.}~\bibnamefont
  {Scardicchio}}\ and\ \bibinfo {author} {\bibfnamefont {T.}~\bibnamefont
  {Thiery}},\ }\href@noop {} {\bibinfo {title} {Perturbation theory approaches
  to anderson and many-body localization: some lecture notes}} (\bibinfo {year}
  {2017}),\ \Eprint {https://arxiv.org/abs/arXiv:1710.01234}
  {arXiv:arXiv:1710.01234 [cond-mat.dis-nn]} \BibitemShut {NoStop}%
\bibitem [{\citenamefont {Gao}\ \emph {et~al.}(2023)\citenamefont {Gao},
  \citenamefont {Khaymovich}, \citenamefont {Wang}, \citenamefont {Xu},
  \citenamefont {Iovan}, \citenamefont {Krishna}, \citenamefont {Balatsky},
  \citenamefont {Zwiller},\ and\ \citenamefont
  {Elshaari}}]{gao2023experimental}%
  \BibitemOpen
  \bibfield  {author} {\bibinfo {author} {\bibfnamefont {J.}~\bibnamefont
  {Gao}}, \bibinfo {author} {\bibfnamefont {I.~M.}\ \bibnamefont {Khaymovich}},
  \bibinfo {author} {\bibfnamefont {X.-W.}\ \bibnamefont {Wang}}, \bibinfo
  {author} {\bibfnamefont {Z.-S.}\ \bibnamefont {Xu}}, \bibinfo {author}
  {\bibfnamefont {A.}~\bibnamefont {Iovan}}, \bibinfo {author} {\bibfnamefont
  {G.}~\bibnamefont {Krishna}}, \bibinfo {author} {\bibfnamefont {A.~V.}\
  \bibnamefont {Balatsky}}, \bibinfo {author} {\bibfnamefont {V.}~\bibnamefont
  {Zwiller}},\ and\ \bibinfo {author} {\bibfnamefont {A.~W.}\ \bibnamefont
  {Elshaari}},\ }\href@noop {} {\bibinfo {title} {Experimental probe of
  multi-mobility edges in quasiperiodic mosaic lattices}} (\bibinfo {year}
  {2023}),\ \Eprint {https://arxiv.org/abs/arXiv:2306.10829}
  {arXiv:arXiv:2306.10829 [cond-mat.dis-nn]} \BibitemShut {NoStop}%
\bibitem [{\citenamefont {Gon\ifmmode~\mbox{\c{c}}\else \c{c}\fi{}alves}\ \emph
  {et~al.}(2023)\citenamefont {Gon\ifmmode~\mbox{\c{c}}\else \c{c}\fi{}alves},
  \citenamefont {Ribeiro},\ and\ \citenamefont
  {Khaymovich}}]{goncalves2023quasiperiodicity}%
  \BibitemOpen
  \bibfield  {author} {\bibinfo {author} {\bibfnamefont {M.}~\bibnamefont
  {Gon\ifmmode~\mbox{\c{c}}\else \c{c}\fi{}alves}}, \bibinfo {author}
  {\bibfnamefont {P.}~\bibnamefont {Ribeiro}},\ and\ \bibinfo {author}
  {\bibfnamefont {I.~M.}\ \bibnamefont {Khaymovich}},\ }\bibfield  {title}
  {\bibinfo {title} {Quasiperiodicity hinders ergodic floquet eigenstates},\
  }\href {https://doi.org/10.1103/PhysRevB.108.104201} {\bibfield  {journal}
  {\bibinfo  {journal} {Phys. Rev. B}\ }\textbf {\bibinfo {volume} {108}},\
  \bibinfo {pages} {104201} (\bibinfo {year} {2023})}\BibitemShut {NoStop}%
\end{thebibliography}%

\clearpage
\begin{widetext}
\appendix
\section{Combined disorder spectra}\label{sec:A1}

\begin{figure*}[ht]
    \centering
    \includegraphics[width=1\textwidth]{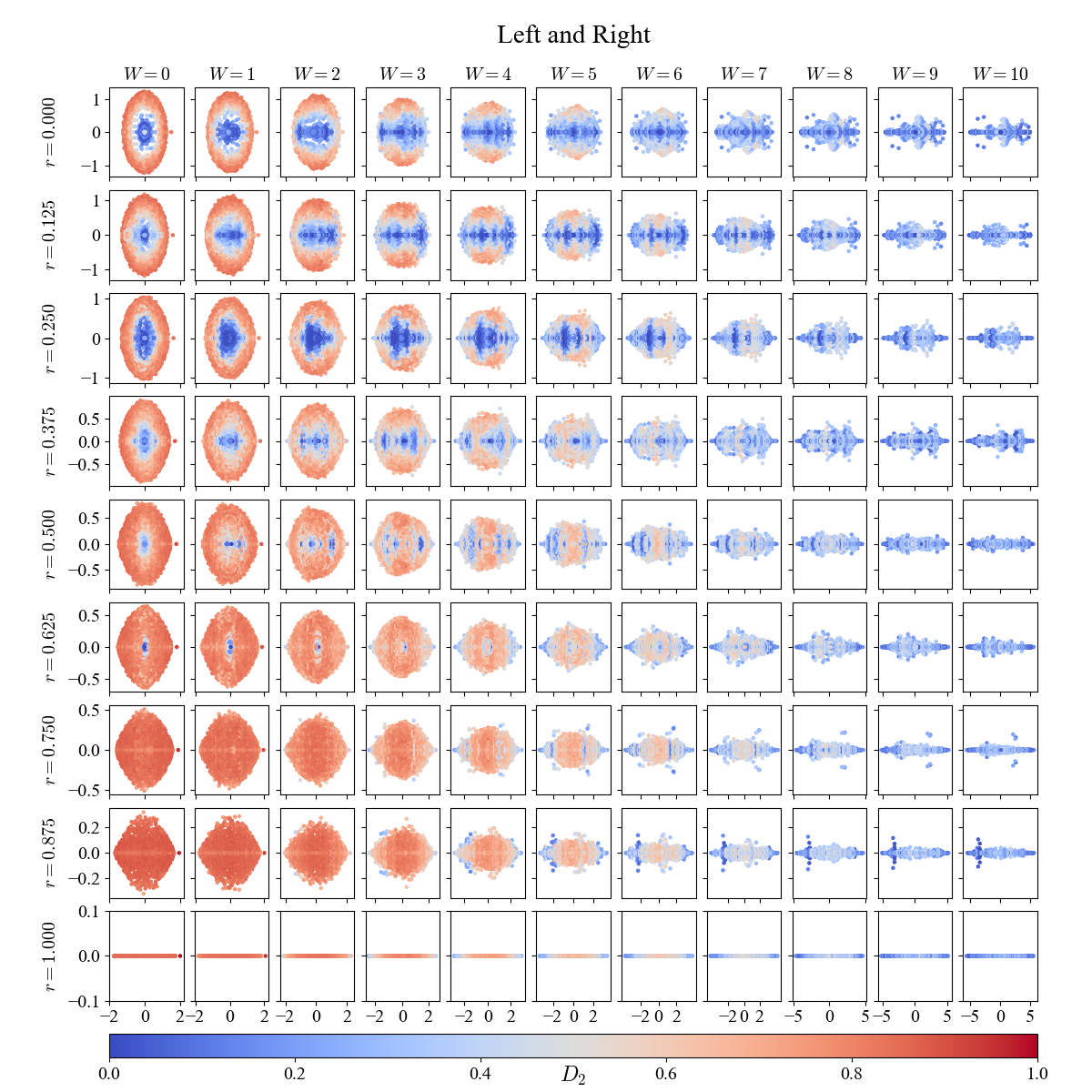}
    \caption{\red{Examples of} spectra of RRG $d=4$ \red{with $N=1024$} adjacency matrices depending on $r$ and $W$, $p=0.5$, colored by right fractal dimensions. For all subfigures, diagonal disorder distribution has the same realization from $[-1/2;+1/2]$ but multiplied by $W$.}
    \label{fig:spec_rW_s}
\end{figure*}

\begin{figure*}[ht]
    \centering
    \includegraphics[width=1\textwidth]{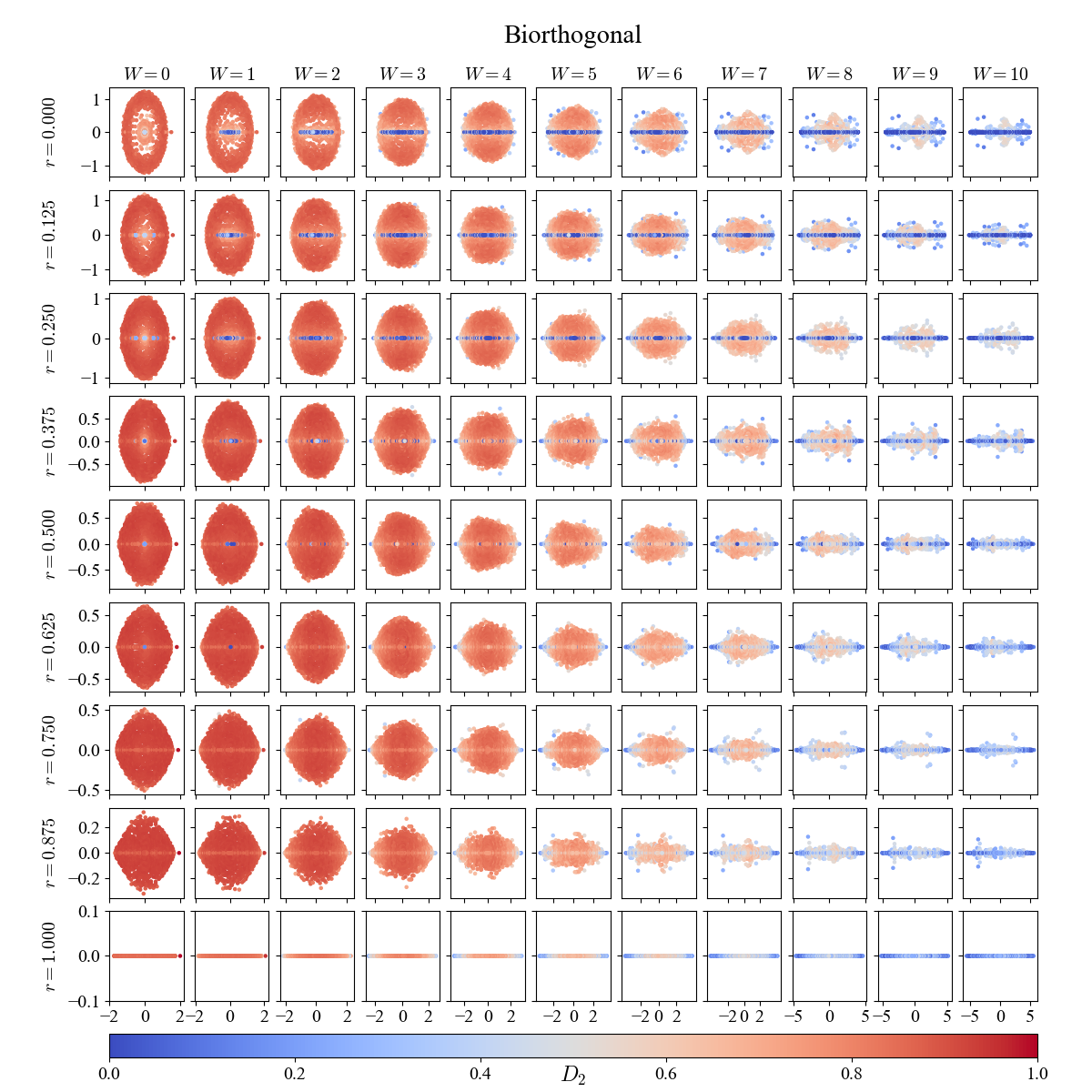}
    \caption{\red{Examples of} spectra of RRG $d=4$ \red{with $N=1024$} adjacency matrices depending on $r$ and $W$, $p=0.5$, colored by biorthogonal fractal dimensions. For all subfigures, diagonal disorder distribution has the same realization from $[-1/2;+1/2]$ but multiplied by $W$.}
    \label{fig:spec_rW_b}
\end{figure*}
\end{widetext}
\end{document}